\documentclass[10pt,twocolumn]{IEEEtran}
\pdfoutput=1
\usepackage{graphicx}
\usepackage{amsmath}
\usepackage{amssymb}
\usepackage[noadjust]{cite}
\usepackage{float}
\usepackage{color}
\usepackage{url}
\usepackage{dblfloatfix} 
\usepackage{xfrac}
\usepackage{amsfonts}
\usepackage{enumitem}
\usepackage{float}
\usepackage{color}
\usepackage{cite}
\usepackage{graphicx}
\usepackage{epstopdf}
\usepackage{bm}
\usepackage{subcaption}
\usepackage{relsize}
\usepackage{mathtools}
\usepackage{array}
\usepackage{array}
\usepackage{multirow}
\usepackage{multicol}
\usepackage{mathtools}
\usepackage{blindtext}
\usepackage{wrapfig}
\usepackage{cuted}
\usepackage{lipsum}
\usepackage{xcolor}
\usepackage{amssymb}
\usepackage{pifont}
\usepackage{makecell}

\newtheorem{Remark}{Remark}

\allowdisplaybreaks

\graphicspath{{Figures/}}
\begin{document}
\title{Altitude-Dependent Cellular Spectrum Occupancy: from Measurements to Stochastic Geometry Models }
\author{Sung Joon Maeng, \.{I}smail G\"{u}ven\c{c},~\IEEEmembership{Fellow, IEEE} \\
\thanks{This work has been supported in part by.}\thanks{\.{I}smail G\"{u}ven\c{c} is with the Department of Electrical and Computer Engineering, North Carolina State University, Raleigh, NC 27606 USA (e-mail: iguvenc@ncsu.edu).} \thanks{Sung Joon Maeng is with the Department of Electrical and Electronic Engineering, Hanyang University, Ansan 15588, South Korea (e-mail: sjmaeng@hanyang.ac.kr)}
}

\maketitle
\begin{abstract}
 The growing demand for aerial connectivity with unmanned aerial vehicles (UAVs) across diverse settings, ranging from urban to rural scenarios, requires developing a better understanding of spectrum occupancy at aerial corridors. In particular, understanding the altitude-dependent behavior of spectrum occupancy in cellular networks, which could be used in the future for enabling beyond visual line of sight (BVLOS) UAV connectivity, is critical. While there are existing models for characterizing altitude-dependent interference in the literature, they are not validated with data and need to be compared with real-world measurements. To address these gaps, in this paper, we conduct cellular spectrum occupancy measurements at various sub-6 GHz bands for altitudes up to 300 meters, in both urban and rural environments. To model the spectrum occupancy measurements, we introduce two different approaches: a theoretical model utilizing stochastic geometry with altitude-dependent factors (SOSGAD), and a ray-tracing model tailored to site-specific line of sight (LOS) and non-LOS scenarios. We analyze the asymptotic behavior of the SOSGAD model as the UAV altitude increases. Through comparative analysis, we assess the effectiveness of the SOSGAD and ray-tracing models for characterizing actual spectrum occupancy as a function of altitude. Our results show that the proposed SOSGAD model can be tuned to closely characterize the real-world spectrum occupancy behavior as the UAV altitude increases. 
\end{abstract}


\section{Introduction}\label{sec:intro}

Due to the flexibility of deployment, scale, and operation, unmanned aerial vehicles (UAVs) have numerous military, public, and commercial applications such as surveillance, search and rescue, drone taxi, and drone delivery. Furthermore, UAVs have promising use cases for advanced wireless communications. For example, UAV-mounted base station (BS) has been studied where UAVs can be used to serve ground users located at the outage region~\cite{xiao2019unmanned,zhang20213d}. Moreover, the UAVs can be utilized as aerial sensor nodes that monitor the spectrum occupancy, which is useful to generate radio emission maps without deploying densely fixed ground sensor nodes~\cite{shang2020spectrum,maeng2024kriging}.

Recently, new services demanding cellular connectivity of UAVs in urban, suburban, and rural scenarios have emerged. Cellular-connected UAVs enable seamless communications and wide-area operations without the vision of a UAV pilot. Advanced air mobility (AAM) and urban air mobility (UAM) will provide brand-new unmanned mobility and transportation services carrying either people or cargo~\cite{namuduri2022advanced,guo2024advanced}. Air corridors are pathways in the sky, which are designed to manage the traffic of a large number of UAVs and AAM~\cite{maeng2024uav}. To support communication services in such scenarios, it is important to understand the spectrum occupancy in aerial corridors considering the altitudes of the UAVs and the AAM traffic. Furthermore, the altitude dependency of line-of-sight (LoS) probability for ground-to-UAV communications is critical, as has been previously investigated in other works in the literature~\cite{al2014modeling,raouf2023spectrum,azari2018uplink}. Thus, it is necessary to characterize the change of spectrum occupancy with altitude.

Stochastic geometry models are widely used in the literature for analyzing the performance of cellular wireless networks where multiple signal sources are spatially distributed with node density by obtaining analytically tractable mathematical expressions~\cite{7733098}. For instance, the analysis of signal-to-interference-plus-noise ratio (SINR) from multiple nodes following uniform spatial node density is useful to evaluate the outage probability of SINR in wireless networks~\cite{rebato2019stochastic,azari2020uav}. Several studies have examined the impact of UAV altitude on the aggregate received signal from spatially distributed ground signal sources~\cite{sinha2021fundamental,banagar2020stochastic,matracia2023uav,ravi2016downlink}. In \cite{azari2018uplink}, stochastic geometry model-based coverage analysis of UE-to-UAV communication network has been studied. However, the paper models the aggregate interference to be independent of the altitude of the UAV, which is not matched with our spectrum occupancy measurements e.g. in~\cite{raouf2023spectrum,maeng2023spectrum}. In ~\cite{sinha2022impact}, the performance of UAV localization from sensor networks on the ground is analyzed considering 3D antenna pattern and altitude-dependent LoS probability. Ray tracing simulations have also been commonly utilized in analyzing the coverage of aerial wireless networks~\cite{al2014modeling,chu2018channel,calvo2018uav}, which can complement measurement studies and theoretical models. The signal power of the source-to-destination is computed based on the 3-D building map and wireless propagation characteristics: signal blockage, diffraction, and reﬂection. 

To our best knowledge, there are no studies in the literature that develop altitude-dependent cellular spectrum occupancy models based on real-world measurements. As such, the validity of the models in earlier works on altitude-dependent interference can not be verified. In this paper, we first present a measurement campaign on altitude-dependent cellular spectrum occupancy in urban and rural environments in sub-6 GHz cellular bands. Subsequently, we introduce three different approaches to model the altitude-dependent spectrum occupancy in these bands. First, we propose a theoretical model which we refer as spectrum occupancy with stochastic geometry and altitude dependence (SOSGAD) for analyzing the aggregate received power of the uplink signal from user equipment (UEs) and the downlink signal from BSs. Second, we analyze the UAV spectrum occupancy by using a ray-tracing model, which enables us to develop site-specific models for LoS/ none-line-of-sight (NLoS) probability based on the real-world environment. Finally, we conduct measurement campaigns to characterize altitude-dependent spectrum occupancy. In the end, we compare the 
 spectrum occupancy versus altitude using the three approaches to understand how closely SOSGAD and ray tracing results are with real-world measurements.

The rest of this paper is organized as follows. Section~\ref{sec:Sys_Model} presents the system model for the UAV spectrum occupancy monitoring scenario. Section~\ref{sec:stoc_geo} introduce a stochastic geometry-based SOSGAD model for spectrum occupancy at a given altitude. In Section~\ref{sec:ray_tracing}, we provide ray-tracing-based analysis results for UAV spectrum occupancy, while in Section~\ref{sec:mea_camp}, we describe the altitude-dependent spectrum measurement campaign using a helikite. In Section~\ref{sec:Results}, we compare the results using the three different approaches, and Section~\ref{sec:Conclusion} provides concluding remarks. 

\section{System Model}\label{sec:Sys_Model}
\begin{figure}[t!]
	\centering
	\includegraphics[width=0.48\textwidth]{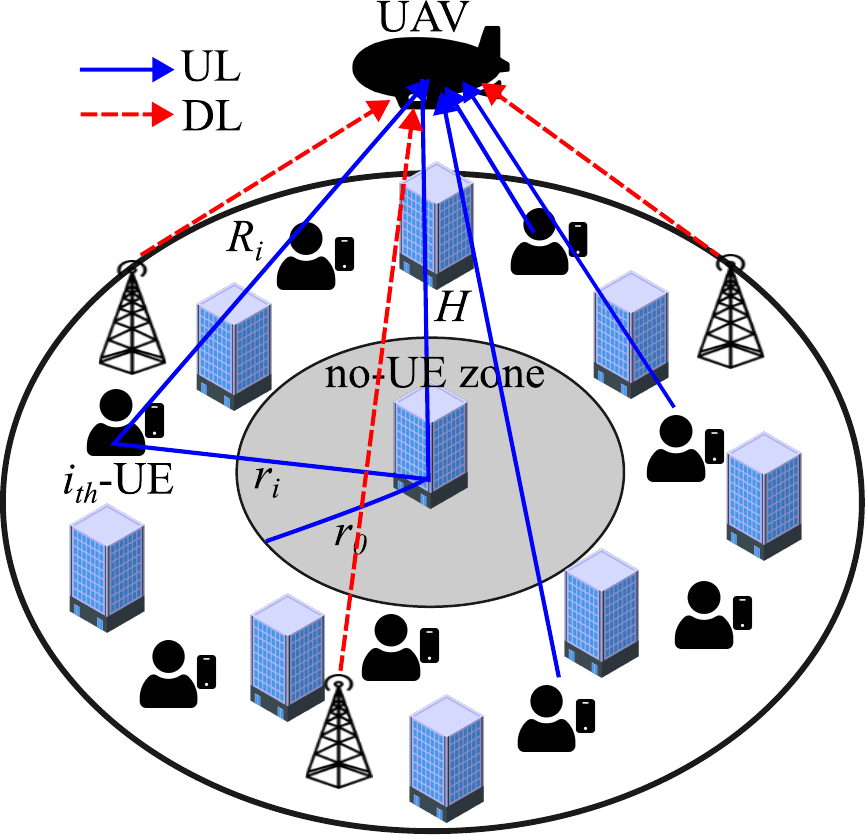}
	\caption{The illustration of the altitude-dependent spectrum occupancy scenario. We conduct spectrum occupancy measurements and develop stochastic geometry based models to characterize the spectrum occupancy both in the downlink and the uplink of cellular networks.}\label{fig:sys_illu}
\end{figure}
We consider a scenario as in Fig.~\ref{fig:sys_illu} where a UAV at a certain altitude is subject to interference from ground signal sources, whether they are UEs (uplink), or eNBs/gNBs (downlink). For the uplink scenario, which is our main focus in this work, we assume the UEs transmit at fixed power $\mathsf{P}_{\rm Tx}$. In addition, it is assumed that the UEs are uniformly deployed with a constant node density $\lambda_{\rm UE}$ following the two-dimensional homogeneous
Poisson point process (HPPP) model. The UAV height in Fig.~\ref{fig:sys_illu} is denoted by ${H}$, and $r_i$, $R_i=\sqrt{r_i^2+{H}^2}$ indicate the vertical distance and the 3D distance between the UAV and $i_{\rm th}$-UE, respectively. We consider a circular no-UE guard zone centered at the helikite with a radius $r_0$, where no UE transmissions are allowed. For the downlink, we consider a similar guard zone and model. While we will focus on the uplink scenario in this section for developing the proposed model, a similar framework is applicable for the downlink model. 

If a UE is located at a visible point from the viewpoint of the UAV, we represent the corresponding wireless link to be in a line-of-sight (LoS) condition. On the other hand, if the direct link between a UE and the UAV is blocked by any obstacles (e.g. buildings), we indicate the link to be in none-line-of-sight (NLoS) condition. Depending on the LoS/NLoS condition, the signal propagation suffers different pathloss, which we will characterize using different models in Section~\ref{sec:rec_sig}.

\section{Proposed SOSGAD-based Analysis}\label{sec:stoc_geo}
In this section, we derive closed-form expressions for the aggregate received signal power of UEs on the ground as observed at the UAV by using stochastic geometry and the LoS probability based models. Furthermore, we analyze the impact of the UAV height, node density, and the radius of the no-UE zone on the received signal power.

\subsection{Spectrum Occupancy Model with Probability LoS}\label{sec:rec_sig}
Considering the HPPP model for the spatial distribution of the UEs on the ground, the aggregate received signal power of the UAV can be expressed as~\cite{7733098,maeng2023analysis} 
\begin{align}\label{eq:rece_0}
    \mathsf{r_{agg}}=\sum_{r_i\in\Phi(\lambda_{\rm UE})\backslash r_0}\frac{\mathsf{P}_{\rm Tx}\mathsf{G}_{\rm t}\mathsf{G}_{\rm r}c^2}{(4\pi)^2f^2R_i^{\alpha}}h_i,
\end{align}
where $h_i\sim \exp(1)$ represents Rayleigh small-scale fading, and $\mathsf{G}_{\rm t}$, $\mathsf{G}_{\rm r}$, $c$, $f$, $\alpha$ denote Tx antenna gain, Rx antenna gain, the speed of light, the carrier frequency, and the pathloss exponent, respectively. We define the spectrum occupancy at a given 3D location as the aggregate received signal power at that location. By using Campbell Theorem~\cite{7733098}, the expectation of the aggregate received signal power over the HPPP can be transformed to
\begin{align}
    \mathbb{E}\left\{\mathsf{r_{agg}}\right\}&=\mathbb{E}\left\{\sum_{r_i\in\Phi(\lambda_{\rm UE})\backslash r_0}\frac{\mathsf{P}_{\rm Tx}\mathsf{G}_{\rm t}\mathsf{G}_{\rm r}c^2}{(4\pi)^2f^2R_i^{\alpha}}h_i\right\},\\
    &=2\pi\lambda_{\rm UE}\frac{\mathsf{P}_{\rm Tx}\mathsf{G}_{\rm t}\mathsf{G}_{\rm r}c^2}{(4\pi)^2f^2}\mathbb{E}\left\{h_i\right\}\int_{R_0}^{\infty} R^{-\alpha+1}{\rm d}R,\\
    &\underset{(a)}{=}2\pi\lambda_{UE}\frac{\mathsf{P}_{\rm Tx}\mathsf{G}_{\rm t}\mathsf{G}_{\rm r}c^2}{(4\pi)^2f^2}\int_{R_0}^{\infty} R^{-\alpha+1}{\rm d}R,\label{eq:rece_1}
\end{align}
where ${R_0}=\sqrt{{H}^2+r_0^2}$, and (a) comes from $\mathbb{E}\left\{h_i\right\}=1$. In the above derivations, we do not take into account the LoS/NLoS conditions of individual links between UEs and the UAV. For now, by applying the path loss exponent for individual link conditions and the LoS probability, we can rewrite \eqref{eq:rece_1} as
\begin{align}
    &\mathbb{E}\left\{\mathsf{r_{agg}}\right\}=2\pi\lambda_{\rm UE}\frac{\mathsf{P}_{\rm Tx}\mathsf{G}_{\rm t}\mathsf{G}_{\rm r}c^2}{(4\pi)^2f^2}\int_{R_0}^{\infty} R^{-\alpha_{\rm los}+1}\mathsf{Pr}_{\rm los}(R){\rm d}R\nonumber\\
    &+2\pi\lambda_{\rm UE}\frac{\mathsf{P}_{\rm Tx}\mathsf{G}_{\rm t}\mathsf{G}_{\rm r}c^2}{(4\pi)^2f^2}\int_{R_0}^{\infty} R^{-\alpha_{\rm Nlos}+1}(1-\mathsf{Pr}_{\rm los}(R)){\rm d}R,
\end{align}
where $\alpha_{\rm los}$, $\alpha_{\rm Nlos}$, $\mathsf{Pr}_{\rm los}(R)$ indicate the pathloss exponent of LoS condition, the pathloss exponent of NLoS condition, and the LoS probability as a function of the 3D distance $R$. Without loss of generality, if we assume that the pathloss exponent of LoS and NLoS are $\alpha_{\rm los}=2$, $\alpha_{\rm Nlos}=3$, then, the aggregate received signal power can be represented as
\begin{align}
    \mathbb{E}\left\{\mathsf{r_{agg}}\right\}&=2\pi\lambda_{\rm UE}\frac{\mathsf{P}_{\rm Tx}\mathsf{G}_{\rm t}\mathsf{G}_{\rm r}c^2}{(4\pi)^2f^2}\left\{\int_{R_0}^{\infty} R^{-1}\mathsf{Pr}_{\rm los}(R){\rm d}R\right.\nonumber\\
    &\left.+\int_{R_0}^{\infty} R^{-2}(1-\mathsf{Pr}_{\rm los}(R)){\rm d}R\right\}.\label{eq:rece_2}
\end{align}
For calculating \eqref{eq:rece_2} in closed form, we need an analytical expression for $\mathsf{Pr}_{\rm los}(R)$ that makes \eqref{eq:rece_2} analytically tractable.

\subsection{Approximation of the Line-of-sight Probability Function}\label{sec:approx_prob}
\begin{figure}[t!]
	\centering
	\subfloat[LoS probability functions in \eqref{eq:PLoS_ori}, \eqref{eq:PLoS_app} depending on the UAV height $H$.]{\includegraphics[width=0.48\textwidth]{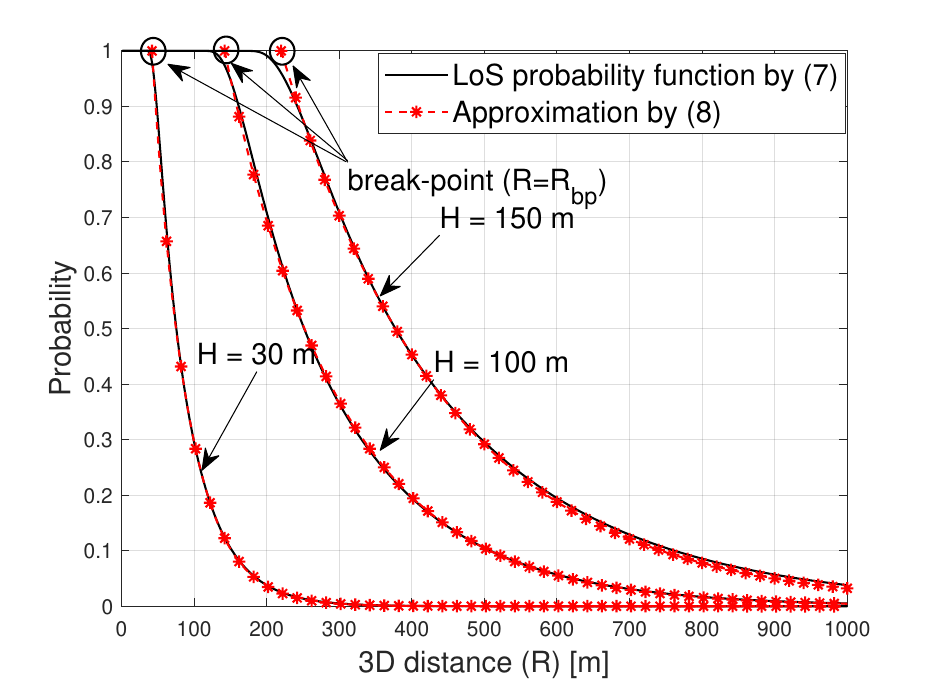}\label{fig:plos_1}}
 
        \subfloat[The values of parameters $k$, $R_{\rm bp}$ with different UAV heights ${H}$. The parameters are fitted by the functions in \eqref{eq:mu_kappa}.]{\includegraphics[width=0.48\textwidth]{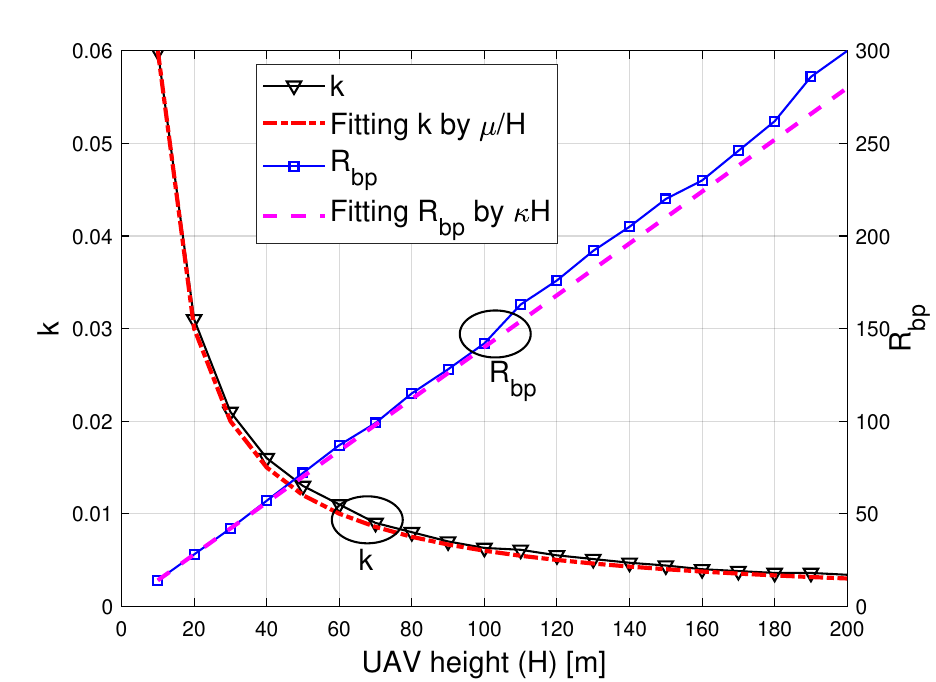}\label{fig:plos_2}}
	\caption{LoS probability function in \eqref{eq:PLoS_ori} with the approximated function by \eqref{eq:PLoS_app}. Parameter fitting results are also plotted.}\label{fig:PLOS}
\end{figure}
It is well-known that the height-dependent LoS probability can be derived as~\cite{6863654}
\begin{align}
    \mathsf{Pr}_{\rm los} = \Pi^{m}_{n=0}\left[1-\exp\left(-\frac{\left({H}-\frac{(n+1/2){H}}{m+1}\right)^2}{2\gamma^2}\right)\right],\label{eq:PLoS_ori}
\end{align}
where $m=\text{floor}(r_i\sqrt{\delta\beta}-1)$, and $\delta$, $\beta$, $\gamma$ are three statistical parameters depending on the environment. In the urban environment, we can use $(\delta, \beta, \gamma)=(0.3, 500, 15)$~\cite{6863654}. Furthermore, the expression in \eqref{eq:PLoS_ori} can be closely approximated to a modified Sigmoid function (S-curve) with respect to the elevation angle $\theta_i=\tan^{-1}\left(\frac{H}{r_i}\right)$~\cite{6863654}. However, the approximated form does not help in obtaining an analytically tractable solution from \eqref{eq:rece_2}.  Alternatively, to facilitate analytical tractability, we introduce a new approximation for the LoS probability expression in \eqref{eq:PLoS_ori} by using a break-point exponential function formula as follows:
\begin{align}\label{eq:PLoS_app}
    \mathsf{Pr}_{\rm los}(R) \approx \begin{cases}
    1 & {H}<R<R_{\rm bp}\nonumber\\
    \exp{(-k(R-R_{\rm bp}))} & R\geq R_{\rm bp}
    \end{cases}~,\\
\end{align}
where $R_{\rm bp}$ denotes a break-point 3D distance where the probability function starts to decrease, and $k$ is a fitting parameter. Note that both $R_{\rm bp}$ and $k$ are functions of the UAV height ${H}$.

Fig.~\ref{fig:plos_1} shows the LoS probability as a function of 3D distance $R$ obtained by the original function in \eqref{eq:PLoS_ori} as well as the approximated version of it in \eqref{eq:PLoS_app}. Note that \eqref{eq:PLoS_ori} can be represented as a function of $R$ given ${H}$ by using the relation $R=\sqrt{r^2+{H}^2}$. It is observed that the approximation by \eqref{eq:PLoS_app} is close to the original function in \eqref{eq:PLoS_ori}. In addition, slight fitting errors can be observed at the break-point in \eqref{eq:PLoS_app}.

In Fig.~\ref{fig:plos_2}, the fitted values of $k$ and $R_{\rm bp}$ in \eqref{eq:PLoS_app} to obtain the results in Fig.~\ref{fig:plos_1} are plotted as a function of the UAV height $H$. From these results, we find that $k$ and $R_{\rm bp}$ can be modeled by inverse and linear functions as follows
\begin{align}
    k&=\frac{\mu}{H},~R_{\rm bp}=\kappa {H}.\label{eq:mu_kappa}
\end{align}
The fitted curves using \eqref{eq:mu_kappa} are also shown in Fig.~\ref{fig:plos_2}. In this result, the values of fitting parameters are $\mu=0.6$, $\kappa=1.38$. Note that by replacing $R_{\rm bp}$, $k$ with \eqref{eq:mu_kappa}, the LoS probability in \eqref{eq:PLoS_app} becomes an explicit function of $R$ and $H$.

\subsection{Closed-from Expressions for SOSGAD}
In this subsection, we derive the closed-form expression for SOSGAD based on the analysis provided in Section~\ref{sec:rec_sig} and Section~\ref{sec:approx_prob}. By plugging in the LoS probability function in \eqref{eq:PLoS_app}, \eqref{eq:mu_kappa} into the aggregate received signal power in \eqref{eq:rece_2}, we can obtain the closed-from expression, as

\begin{align}\label{eq:rece_3}
    \mathbb{E}\left\{\mathsf{r_{agg}}\right\}=&2\pi\lambda_{\rm UE}\frac{\mathsf{P}_{\rm Tx}\mathsf{G}_{\rm t}\mathsf{G}_{\rm r}c^2}{(4\pi)^2f^2}\left\{\int_{R_0}^{R_{\rm bp}} R^{-1}{\rm d}R\right.\nonumber\\
    &\left.+\int_{R_{\rm bp}}^{\infty} \frac{\exp{(-k(R-R_{\rm bp}))}}{R}{\rm d}R\right.\nonumber\\
    &\left.+\int_{R_{\rm bp}}^{\infty} \frac{1-\exp{(-k(R-R_{\rm bp}))}}{R^2}{\rm d}R\right\}\nonumber\\
    &=2\pi\lambda_{\rm UE}\frac{\mathsf{P}_{\rm Tx}\mathsf{G}_{\rm t}\mathsf{G}_{\rm r}c^2}{(4\pi)^2f^2}\left\{\log\left(\frac{R_{\rm bp}}{R_0}\right)\right.\nonumber\\
    &\left.+\exp{(kR_{\rm bp})}\Gamma(0,kR_{\rm bp})\right.\nonumber\\
    &\left.+\frac{1}{R_{\rm bp}}-k\exp{(kR_{\rm bp})}\Gamma(-1,kR_{\rm bp})\right\},
\end{align}
where $\Gamma(s,x)=\int_{x}^{\infty}R^{s-1}\exp{(-R)}{\rm d}R$ denotes the upper incomplete gamma function. Then, we can plug \eqref{eq:mu_kappa} into \eqref{eq:rece_3} to incorporate altitude dependence for a specific height. The closed-form expression of the spectrum occpancy can then be written as 
\begin{align}\label{eq:rece_4}
    \mathbb{E}\left\{\mathsf{r_{agg}}\right\}&=2\pi\lambda_{\rm UE}\frac{\mathsf{P}_{\rm Tx}\mathsf{G}_{\rm t}\mathsf{G}_{\rm r}c^2}{(4\pi)^2f^2}\left\{\log\left(\frac{\kappa {H}}{R_0}\right)\right.\nonumber\\
    &\left.+\exp{(\mu\kappa)}\Gamma(0,\mu\kappa)\right.\nonumber\\
    &\left.+\frac{1}{\kappa {H}}-\frac{\mu}{H}\exp{(\mu\kappa)}\Gamma(-1,\mu\kappa)\right\}.
\end{align}

From the closed-form expression in \eqref{eq:rece_4}, we can analyze the asymptotic behavior of the received signal power as the UAV height $H$ increases.
\newtheorem{theorem}{Theorem}
\begin{theorem}\label{t1}
The average received signal power converges to
\begin{align}\label{eq:rece_asym}
    &\mathbb{E}\left\{\mathsf{r_{agg}}\right\}\to\nonumber\\
    &2\pi\lambda_{\rm UE}\frac{\mathsf{P}_{\rm Tx}\mathsf{G}_{\rm t}\mathsf{G}_{\rm r}c^2}{(4\pi)^2f^2}\left\{\log\left(\kappa\right)+\exp{(\mu\kappa)}\Gamma(0,\mu\kappa)\right\},\;\text{as}\;{H}\to\infty.
\end{align}
\end{theorem}
\begin{IEEEproof}
We can easily obtain \eqref{eq:rece_asym} from \eqref{eq:rece_4} by using $\frac{1}{H}\to 0$, $R_0\to{H},\;\text{as}\;{H}\to\infty$. 
\end{IEEEproof}

\begin{Remark}\label{r1}
It is important to highlight the impact of the height of a UAV on the received signal power. When the height of a UAV increases, 1) the received signal power from an individual UE decreases due to the increasing pathloss at a larger distance; 2) simultaneously, the number of UEs connected by the LoS condition increases (converted from NLoS to LoS), which increases the aggregate received signal power. Interestingly, \textit{Theorem~1} implies that combining two factors related to the received signal power asymptotically compensates for each other and results in convergence for large $H$. Note that while a related observation appears in \cite{azari2018uplink}, the aggregate received power is modeled to be constant even for the low height.
\end{Remark}

\subsection{Behavior of SOSGAD with $r_0$, $\lambda_{\rm UE}$}\label{sec:lammda_r0}
We first consider the case of $r_0=0$ where the UEs can be closely located near the UAV. When the radius of no-UE zone $r_0 = 0$, the average received signal power in \eqref{eq:rece_4} becomes
\begin{align}\label{eq:ragg_r0_1}
    \mathbb{E}\left\{\mathsf{r_{agg}}\right\}&=2\pi\lambda_{\rm UE}\frac{\mathsf{P}_{\rm Tx}\mathsf{G}_{\rm t}\mathsf{G}_{\rm r}c^2}{(4\pi)^2f^2}\left\{\log\left(\kappa\right)+\exp{(\mu\kappa)}\Gamma(0,\mu\kappa)\right.\nonumber\\
    &\left.+\left(\frac{1}{\kappa}-\mu\exp{(\mu\kappa)}\Gamma(-1,\mu\kappa)\right)\frac{1}{{H}}\right\}.
\end{align}
Since the coefficient of the third term $\frac{1}{\kappa}-\mu\exp{(\mu\kappa)}\Gamma(-1,\mu\kappa)$ is a positive real-value, the received signal power in \eqref{eq:ragg_r0_1} gradually decreases as ${H}$ increases.

Next, for the case that $r_0$ is significantly larger than ${H}$ $(r_0\gg{H})$, we can replace $R_0 \to r_0$ in \eqref{eq:rece_4}. Then, the average received signal power becomes
\begin{align}\label{eq:ragg_r0_2}
    &\mathbb{E}\left\{\mathsf{r_{agg}}\right\}\to2\pi\lambda_{\rm UE}\frac{\mathsf{P}_{\rm Tx}\mathsf{G}_{\rm t}\mathsf{G}_{\rm r}c^2}{(4\pi)^2f^2}\left\{\log({H})+\log\left(\frac{\kappa}{r_0}\right)\right.\nonumber\\
     &\left.+\exp{(\mu\kappa)}\Gamma(0,\mu\kappa)+\left(\frac{1}{\kappa}-\mu\exp{(\mu\kappa)}\Gamma(-1,\mu\kappa)\right)\frac{1}{{H}}\right\}.
\end{align}
In \eqref{eq:ragg_r0_2}, the first term $\log({H})$ is dominant, and the received signal power gradually increases as the UAV height increases.

\begin{Remark}\label{r2}
The aggregate received signal power can be either an increasing or decreasing function of $H$ depending on the value of $r_0$. When $r_0$ is sufficiently small, the average received signal power gradually decreases and converges to the value in \eqref{eq:rece_asym}, while when $r_0$ becomes large, the average received signal power gradually increases and converges to the value in \eqref{eq:rece_asym}. This implies that UEs located sufficiently close to the UAV can make the received signal power a decreasing function of the UAV height. Otherwise, the received signal power is an increasing function of the UAV height. In addition, the no-UE zone can be interpreted as a result of the UAV's antenna pattern, which may create an outage region directly below the UAV, considering the characteristics of a typical dipole antenna pattern. Note that regardless of $r_0$, the received signal power converges to the same value from \textit{Theorem~1}.
\end{Remark}

\begin{Remark}\label{r3}
From the closed-form expression \eqref{eq:rece_4}, we can also observe that the UE node density $\lambda_{\rm UE}$ is linearly proportional to the average received signal power $\mathbb{E}\left\{\mathsf{r_{agg}}\right\}$. This implies that a growing number of UEs in the same area increases the received signal power of the UAV.
\end{Remark}

\section{Ray-tracing-based Spectrum Occupancy Analysis with Altitude Dependence}\label{sec:ray_tracing}
In this section, we introduce a framework for ray-tracing-based spectrum occupancy analysis with altitude dependence. We create a grid of the UEs' locations on the 3-D map and determine the LoS/NLoS link status based on the ray-tracing analysis carried out in a specific environment. At the end, we obtain the aggregate received signal power as a function of the UAV height.

\subsection{Visualization of Real-world 3-D Map}
\begin{figure}[t!]
	\centering
	\subfloat[OSM data with 3D visualization using Site Viewer.]{\includegraphics[width=0.45\textwidth]{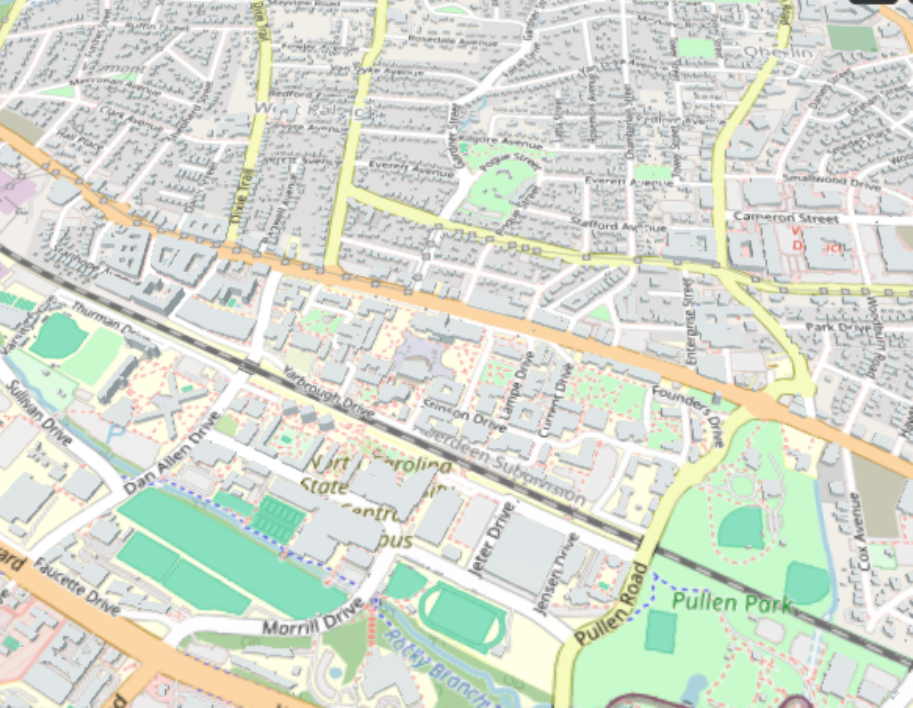}\label{fig:OSM_1}}\vspace{2mm}
 
        \subfloat[An example of LoS/NLoS condition analysis by using ray-tracing model, with a UAV at $H=30$~m.]{\includegraphics[width=0.45\textwidth]{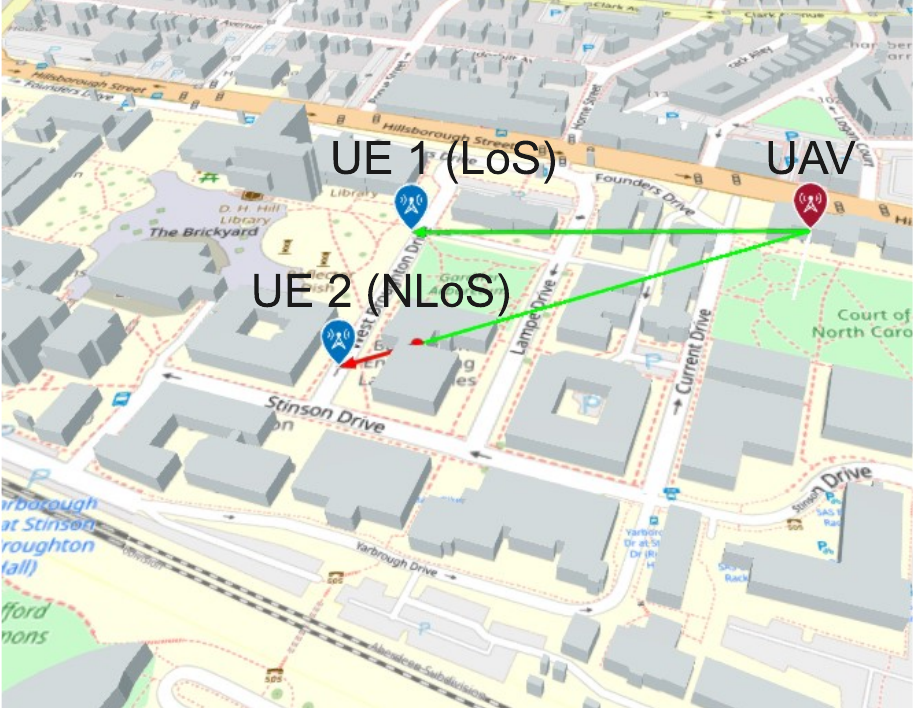}\label{fig:OSM_2}}
	\caption{Map construction using by OSM data where 3-D buildings are integrated into the map. LoS/NLoS conditions are analyzed using ray-tracing.}\label{fig:OSM}
\end{figure}
\begin{figure}[t!]
	\centering
	\subfloat[UAV height (H) = 5~m.]{\includegraphics[width=0.22\textwidth]{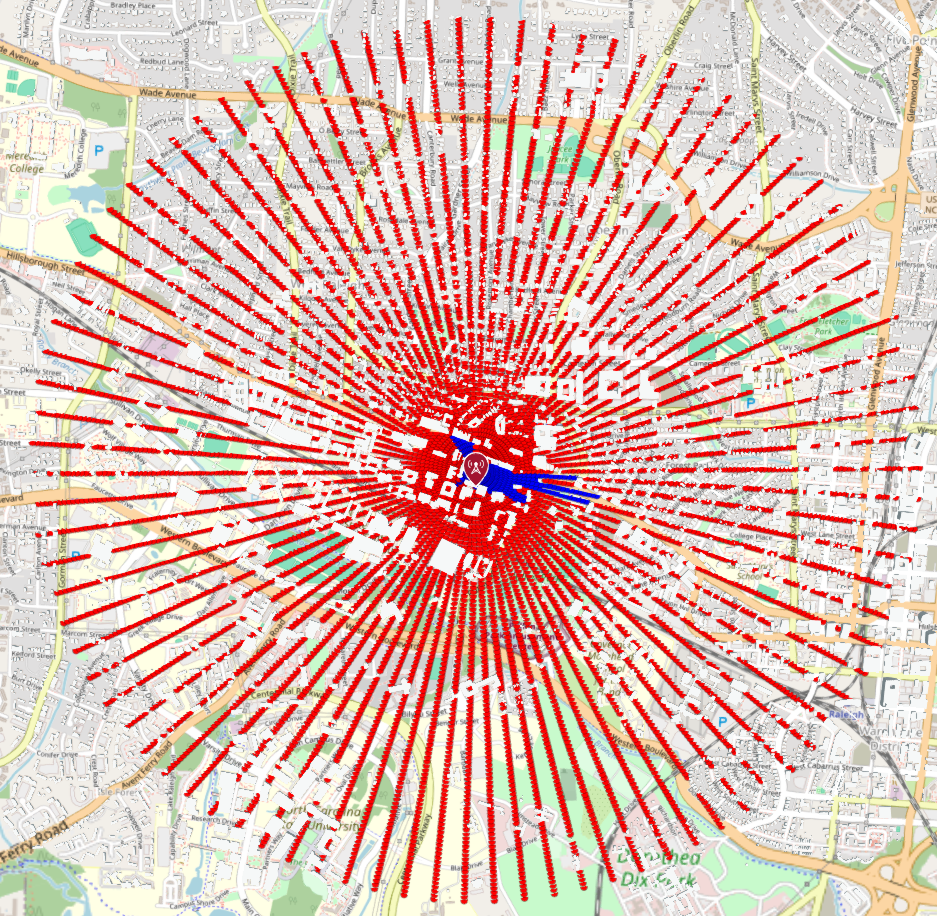}}~
        \subfloat[UAV height (H) = 30~m.]{\includegraphics[width=0.22\textwidth]{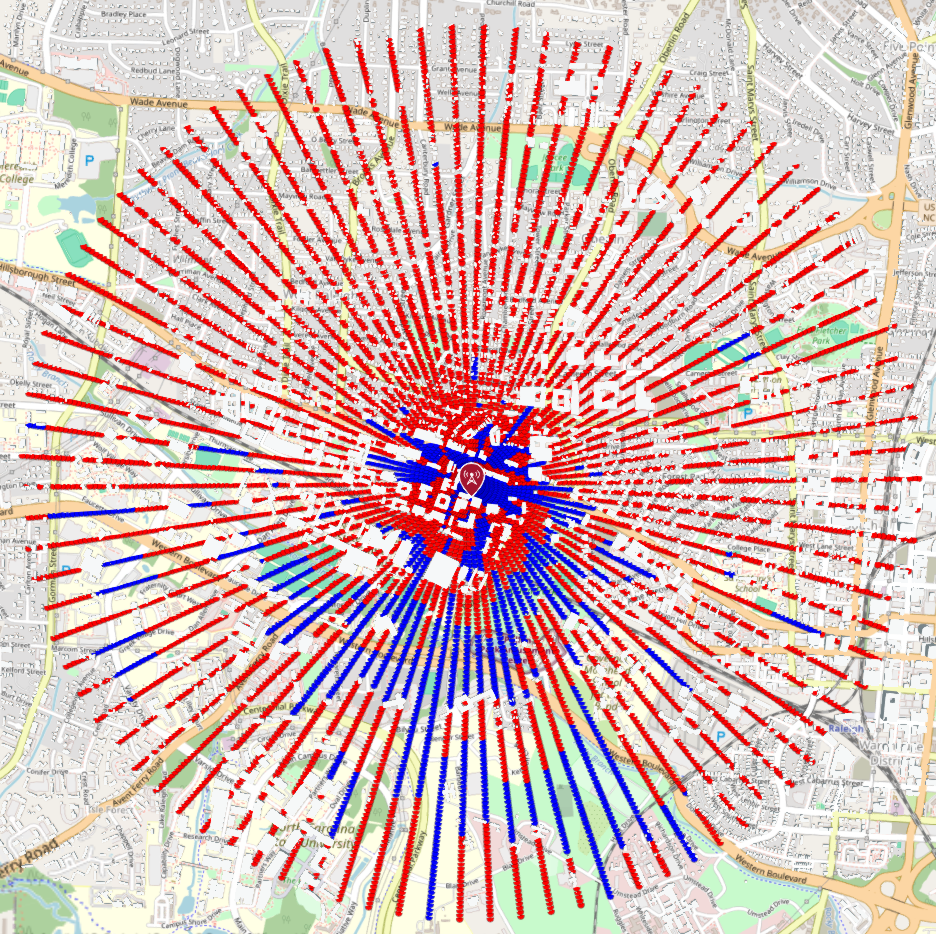}}

        \subfloat[UAV height (H) = 60~m].{\includegraphics[width=0.22\textwidth]{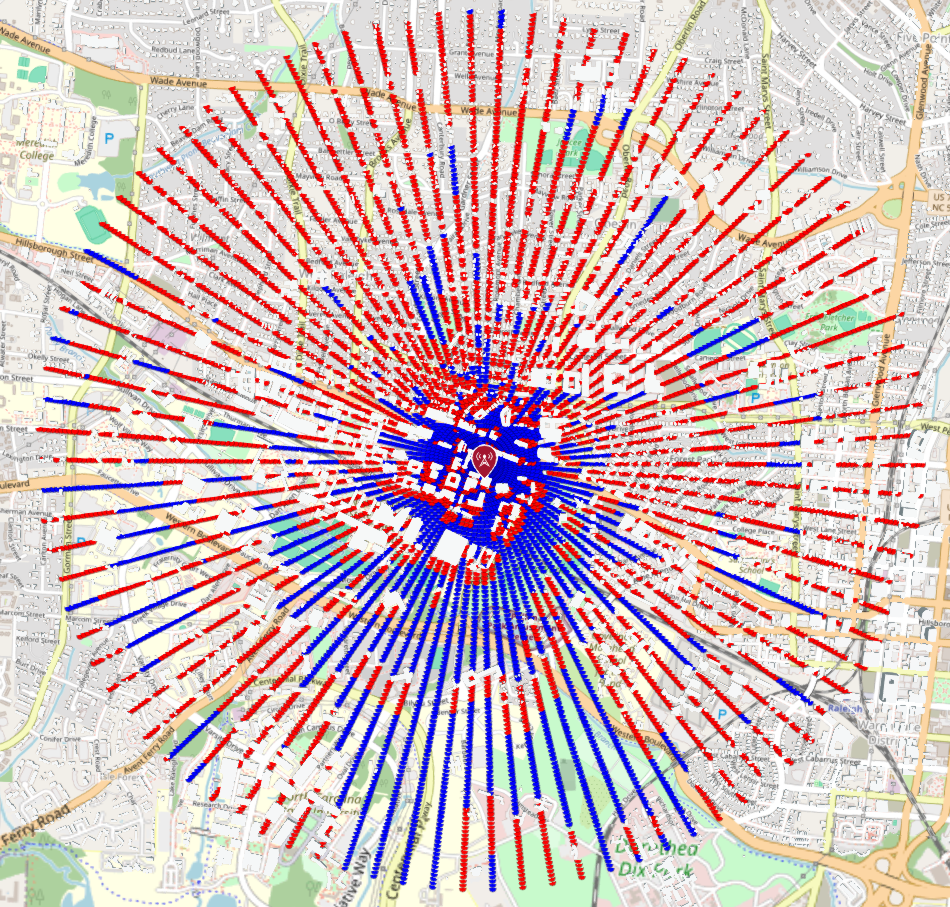}}~
        \subfloat[UAV height (H) = 100~m.]{\includegraphics[width=0.22\textwidth]{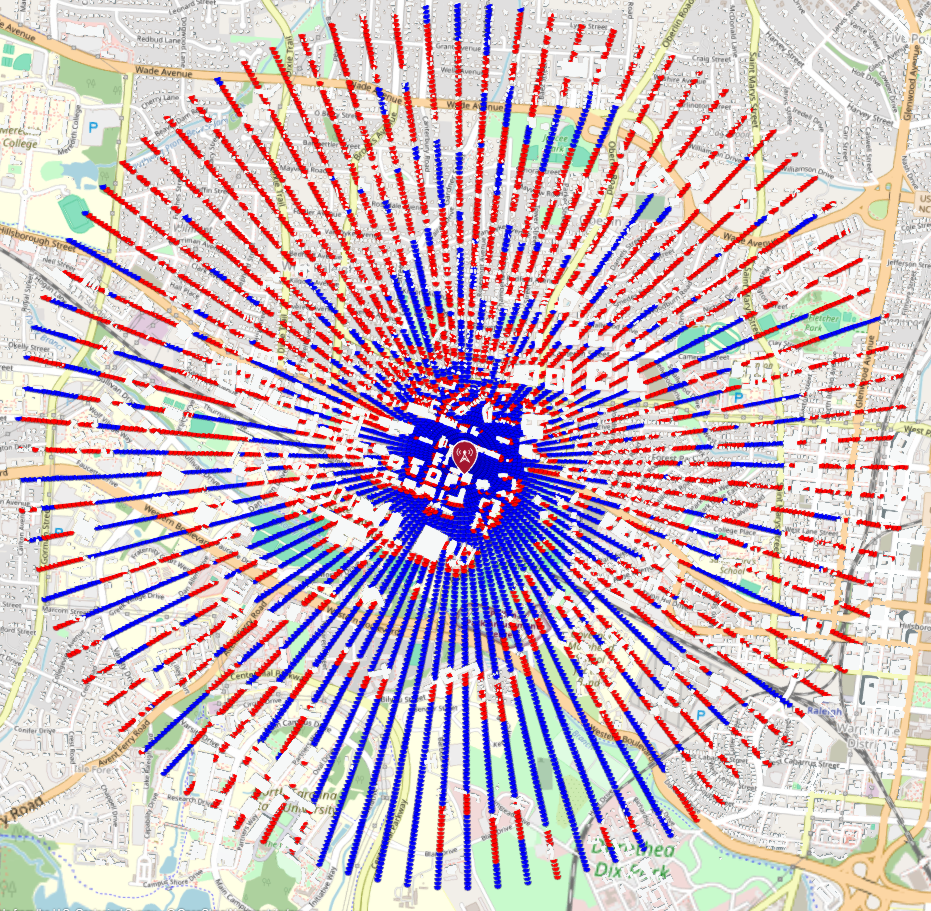}}

        \subfloat[UAV height (H) = 200~m.]{\includegraphics[width=0.22\textwidth]{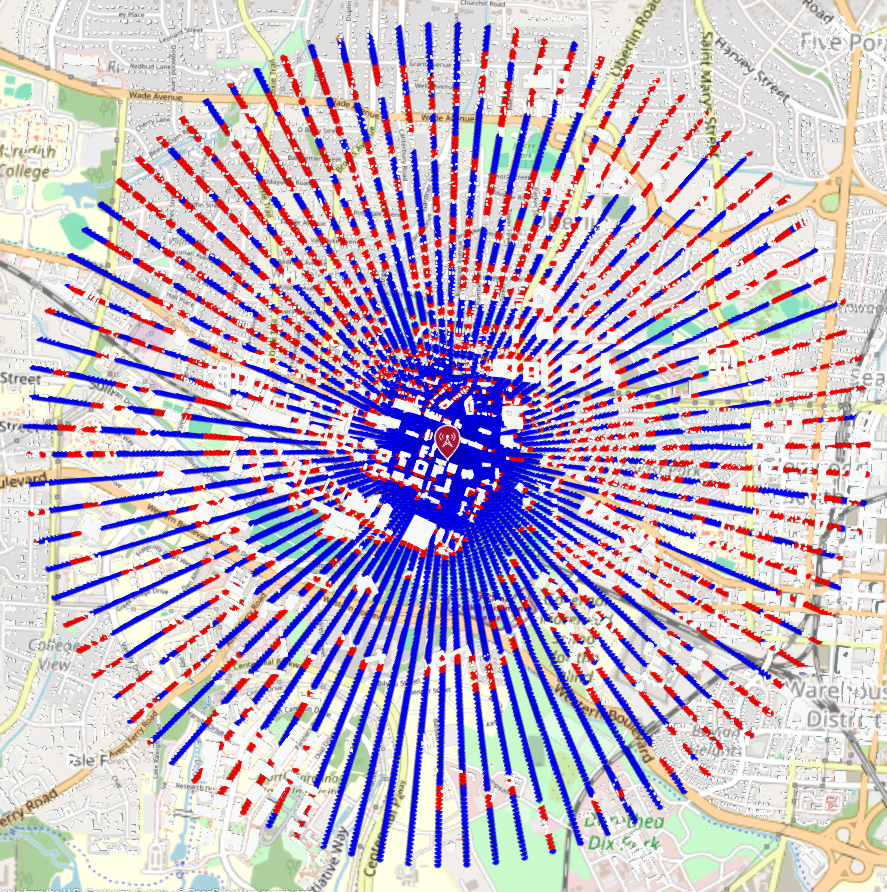}}~
        \subfloat[UAV height (H) = 300~m.]{\includegraphics[width=0.22\textwidth]{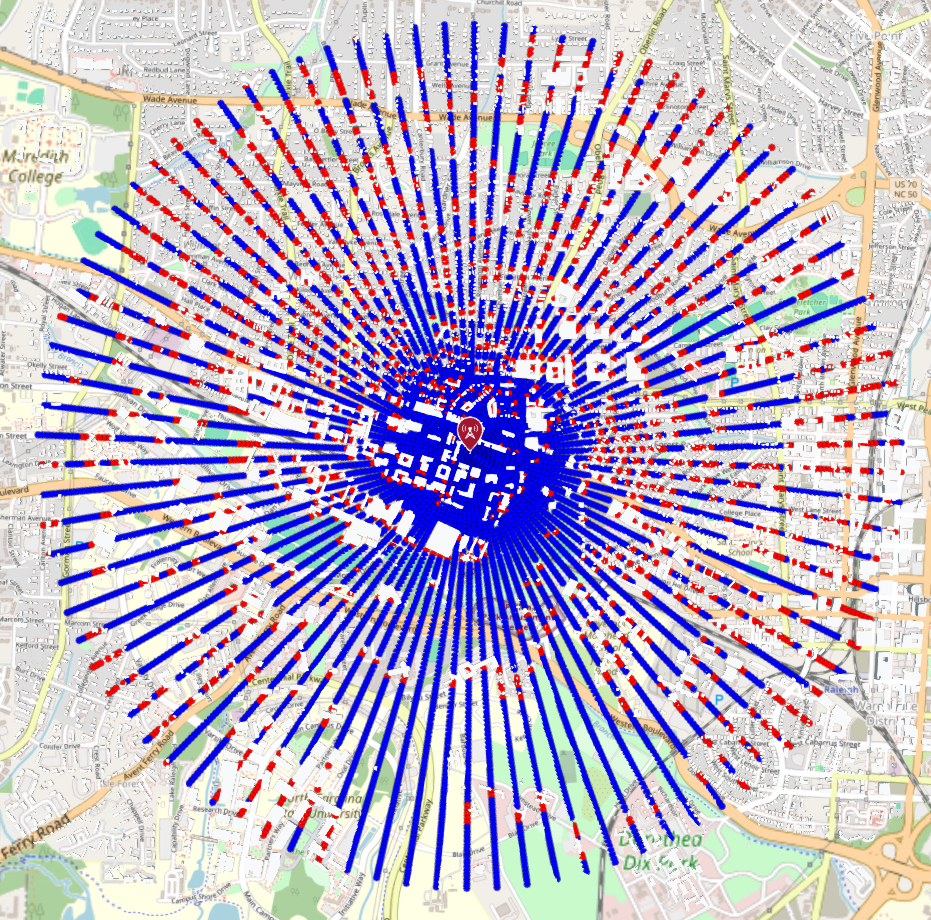}}
	\caption{LoS/NLoS condition analysis based on ray-tracing depending on the UAV height where the color blue(red) indicates LoS(NLoS) links. The UAV is located at the center.}\label{fig:LoS_ray}
\end{figure}
We download the real-world geographical map data from OpenStreetMap (OSM) for the considered spectrum measurement scenario~\cite{OSM}. We utilize QGIS and JOSM programs to extract the maps and buildings data of targeted regions from OSM. The OSM data includes information on 3-D building shape and building height. To visualize the OSM data, we utilize Site Viewer in MATLAB Antenna Toolbox. Fig.~\ref{fig:OSM_1} shows the visualized 3-D map obtained by importing the OSM data. The visualized location of the map includes the region of the Main Campus of North Carolina State University in Raleigh NC, USA, where the spectrum occupancy measurement experiments are conducted. We can see that the 3-D rendered buildings are integrated into the map. 

\subsection{LoS Visibility Analysis Based on Ray-tracing Model}
In MATLAB Site Viewer environment, by setting the locations (latitude, longitude, antenna height) of the transmitter and the receiver over the 3-D map data, we can identify the visibility of LoS between any Tx and Rx location by using ray-tracing analysis. Fig.~\ref{fig:OSM_2} shows an example of the ray-tracing LoS visibility analysis results with two different locations of UEs. It is shown that UE-1 is visible from the viewpoint of the UAV, while UE-2 is blocked by a building. From this analysis, we can determine the LoS condition of UE-1 and the NLoS condition of UE-2 for this specific environment and UAV location. Note that we only utilize the ray-tracing model to analyze the LoS/NLoS status, and the received signal power is not directly obtained by ray-tracing. Instead, different path loss models will be used for LoS and NLoS links to obtain the aggregate signal, as discussed in the next section.

\subsection{UE Deployment and Received Signal Power}
We deploy the UEs on the 3-D map by forming a circular grid with a maximum radius of $2$~km centered at the location of the UAV as shown in Fig.~\ref{fig:LoS_ray}. The location of the UAV is considered to be the same location of the UAV in the field experiment that will be discussed in Section~\ref{sec:mea_camp}. 
The resolution of the radius and the azimuth angle for the circular grid are set to $20$~m and $4$~degrees, respectively. Therefore, the total number of UEs on the grid is ${\rm N}_{\rm UE}=100\times90=9000$. Note that we use a single grid resolution setup in this ray-tracing simulation, which implies a fixed spatial density of UEs. When the locations of UEs are overlapped with buildings, we remove the UEs in the analysis. It is observed in Fig.~\ref{fig:LoS_ray} that the NLoS links of UEs (red) convert into LoS links (blue) as the UAV height (${H}$) increases.

After all the LoS/NLoS conditions of UEs are obtained by ray-tracing analysis, by using \eqref{eq:rece_0}, we calculate the received signal power of UE-$i$ as
\begin{align}
    \mathsf{r}_i&=
    \begin{cases}\label{eq:}
    \frac{\mathsf{P}_{\rm Tx}\mathsf{G}_{\rm t}\mathsf{G}_{\rm r}c^2}{(4\pi)^2f^2R_i^{\alpha_{\rm los}}}h_i & \text{: UE-$i$ is in LoS condition}\\
    \frac{\mathsf{P}_{\rm Tx}\mathsf{G}_{\rm t}\mathsf{G}_{\rm r}c^2}{(4\pi)^2f^2R_i^{\alpha_{\rm Nlos}}}h_i & \text{: UE-$i$ is in NLoS condition}
    \end{cases}~,
\end{align}
with the total power from all UEs given by
\begin{align}
    \mathsf{r_{agg}}&=\sum_{i=1}^{{\rm N}_{\rm UE}} \mathsf{r}_i.
\end{align}
Note that the distance $R$ between UEs and the UAV can be calculated by the 3-D coordinates (latitude, longitude, antenna height).

\section{Altitude-Dependent Spectrum Occupancy Measurement Campaign}\label{sec:mea_camp}
\begin{figure}[t!]
	\centering
        \subfloat[Main Campus of NC State University, with downtown Raleigh in the background (urban).]{\includegraphics[width=0.45\textwidth]{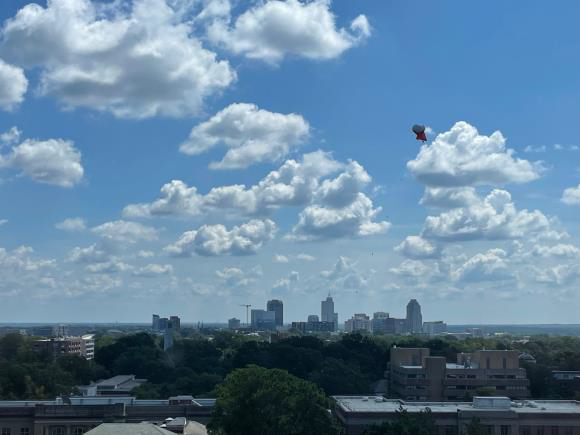}\label{fig:heli_exp}}\vspace{2mm}
 
        \subfloat[LakeWheeler field lab (rural).]{\includegraphics[width=0.45\textwidth]{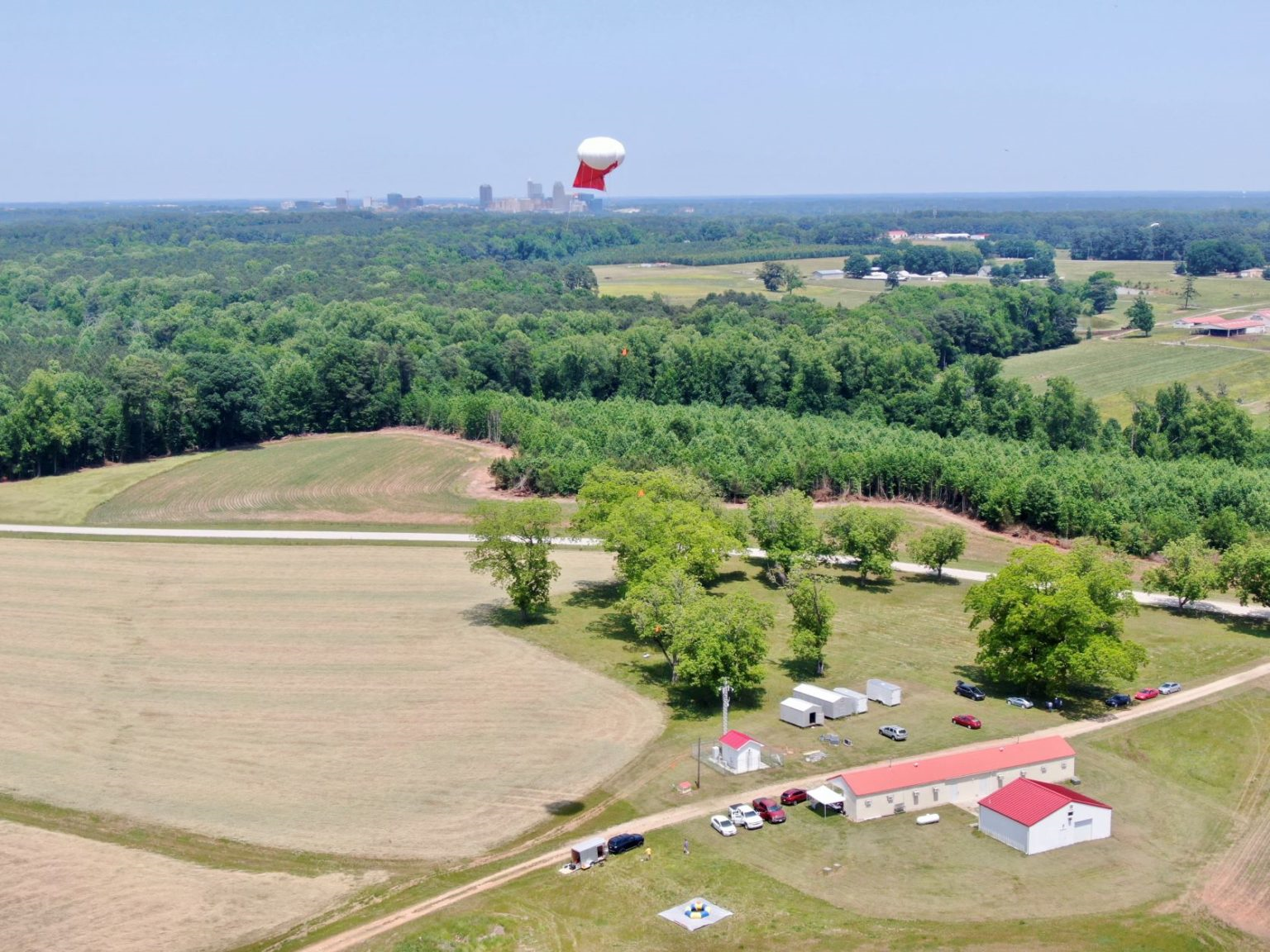}\label{fig:heli_exp_lw}}
	\caption{Photo of helikite flying over the experiment sites.}
\end{figure}
\begin{figure}[t!]
	\centering
        \includegraphics[width=0.48\textwidth]{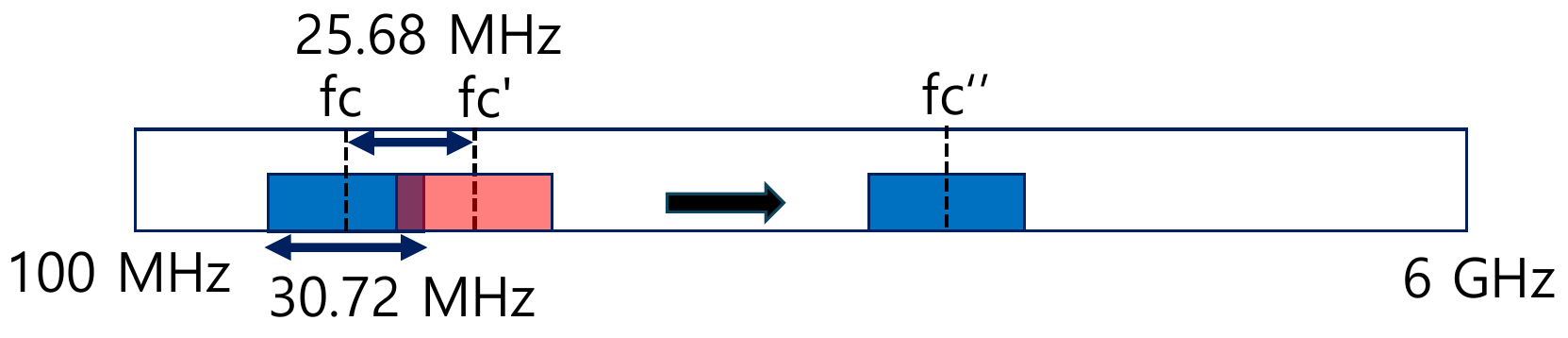}
	\caption{Spectrum sweep procedure of experiments where sampling rate is 30.72~MHz and center frequency shift is 25.68~MHz.}\label{fig:spec_moni}
\end{figure}
In this section, we describe a spectrum occupancy monitoring experiment where a helikite is used to collect data in rural and urban areas. The experiments was conducted at the Main Campus of NC State University during the Packapalooza Festival in August 2023. A photo of the helikite during the experiment is shown in Fig.~\ref{fig:heli_exp}. Since the helikite is available to float for a long duration without external power, it was possible to collect large amount of spectrum occupancy data at different altitudes. A similar experiment was conducted in the rural area at LakeWheeler field lab in Aug 2024 in Fig.~\ref{fig:heli_exp_lw} in order to compare spectrum occupancy in rural and urban environments. Note that while the rural measurement campaign was a more controlled experiment exclusively designed for spectrum occupancy measurements, the urban experiment was done while the helikite was used also by festival organizers for security purposes; hence it was not possible to fully control its altitude over time.

Datasets obtained from these spectrum measurements as well as the post-processing scripts to obtain the results in this manuscript can be found at~\cite{dataset}.

\subsection{Helikite Experiment Setup}
\begin{figure*}[t!]
	\centering
	\subfloat[3-D trajectory of the helikite during the urban experiment.]{\includegraphics[width=0.24\textwidth]{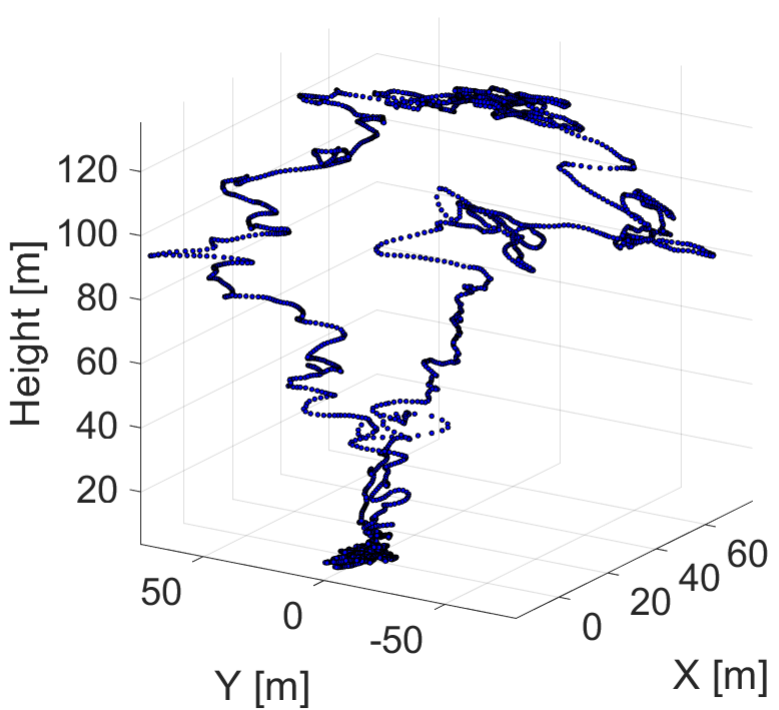}\label{fig:heli_traj}}~
 	\subfloat[3-D trajectory of the helikite during the rural experiment.]{\includegraphics[width=0.24\textwidth]{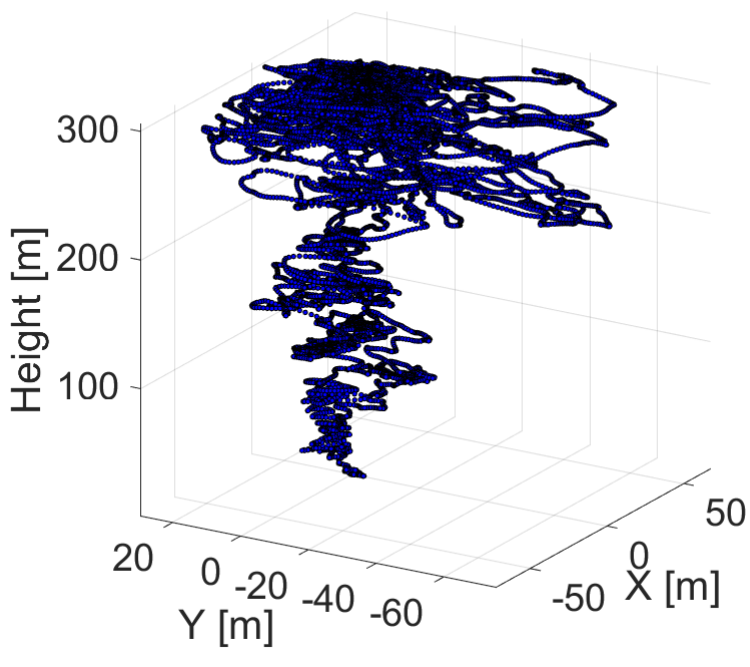}\label{fig:heli_traj_lw}}~
        \subfloat[Helikite height during the urban experiment.]{\includegraphics[width=0.24\textwidth]{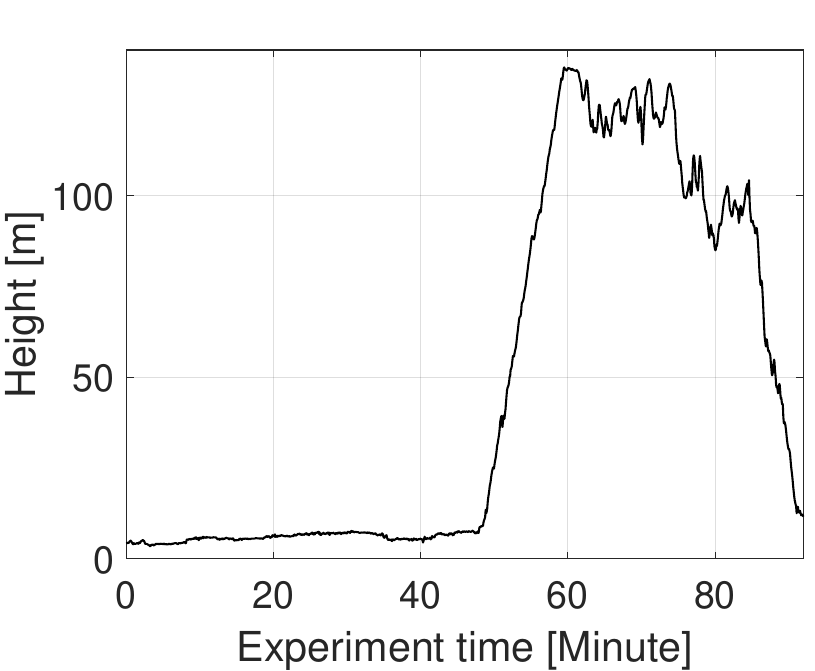}\label{fig:heli_height}}~
        \subfloat[Helikite height during the rural experiment.]{\includegraphics[width=0.24\textwidth]{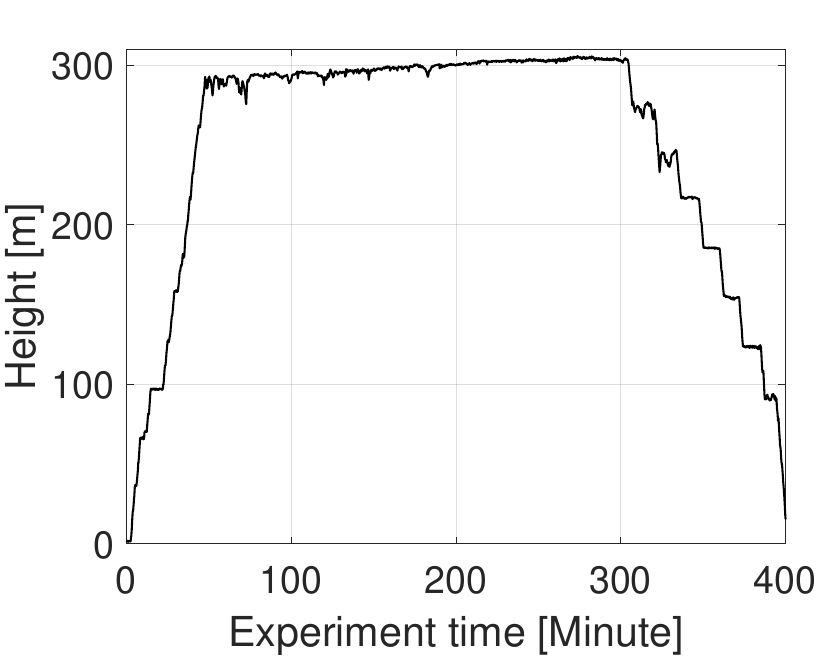}\label{fig:heli_height_lw}}
	\caption{The trajectory and height of the helikite during the urban and rural experiments as obtained from the GPS logs.}\label{fig:heli_GPS}
\end{figure*}
\begin{table}[!t]
\caption{List of frequency bands considered for the United States LTE / NR networks~\cite{listofnetworks}.}
\label{table:LTE_bands}
\centering
\begin{tabular}{p{1.8cm}|p{1.0cm}|p{2.3cm}|p{2cm}}
\hline
Band No & Duplex Mode & Uplink Band (MHz) & Operators \\
\hline\hline
n71 (5G) & \centering FDD & \centering 663 - 698 & T-Mobile \\
12 (LTE)& \centering FDD & \centering698 - 716 & AT\&T, T-Mobile \\
5, n5 (LTE, 5G)\textbf{} &\centering FDD & \centering824 - 849 & AT\&T, T-Mobile, Verizon \\
4 (LTE)& \centering FDD & \centering1710 - 1755 & AT\&T, T-Mobile, Verizon \\
2, n2 (LTE, 5G)& \centering FDD & \centering1850 - 1910 & AT\&T, T-Mobile, Verizon \\
30 (LTE)& \centering FDD & \centering2305 - 2315 & AT\&T \\
\hline\hline
\end{tabular}
\end{table}
For the measurement experiments, a software-defined radio (SDR) and a global positioning system (GPS) receiver were mounted on the helikite. The helikite-mounted SDR continuously measured the spectrum from $87$~MHz to $6$~GHz. The summary of the spectrum sweep procedure is provided in Fig.~\ref{fig:spec_moni}. 
 In the urban area experiment, a single sweep took around 15 seconds and a total of 364 sweeps were conducted during the experiment. The GPS receiver continuously recorded the 3-D position of the helikite. Fig.~\ref{fig:heli_traj} and Fig.~\ref{fig:heli_height} show the trajectory and the height of the helikite during the experiment where the location information is obtained from the GPS logs. Depending on the wind power and direction, the helikite randomly drafts in the air. We adjusted the height of the helikite by controlling the length of the tether tied to the helikite. The experiment took around 90 minutes, and the height of the helikite stayed around 5~m for half of the experiment, and a height of 100~m to 140~m for the majority of the remaining time. At the end of the experiment, the helikite gradually went down to the ground. By post-processing the GPS logs and spectrum measurements dataset, we generate the integrated dataset of time, position, frequency, and received signal power~\cite{dataset}. In the rural area experiment, a total of 1524 sweeps were conducted for around 400 minutes. Fig.~\ref{fig:heli_traj_lw} and Fig.~\ref{fig:heli_height_lw} show the trajectory and the height of the helikite during the experiment. The helikite went up to a height of around 300~m and stayed around 4 hours and went down to the ground. For these rural experiments, permission is received from FAA to fly the helikite up to 300 m -- this data is especially valuable to understand the trends in spectrum occupancy at higher altitudes.

In this work, we focus on utilizing the spectrum of the specific uplink bands of LTE and 5G NR as listed in Table~\ref{table:LTE_bands}. The comprehensive results of spectrum occupancy monitoring measurement for the whole sweeping bands can be found in~\cite{raouf2023spectrum}.

\section{Numerical Results}\label{sec:Results}
In this section, we present spectrum occupancy results for different scenarios, considering the SOSGAD model introduced in Section~\ref{sec:stoc_geo}, ray tracing based analysis discussed in Section~\ref{sec:ray_tracing}, and the measurements spectrum occupancy measurments described in Section~\ref{sec:mea_camp}.

\subsection{Spectrum Occupancy Results from SOSGAD Model}
\begin{table}[!t]
\renewcommand{\arraystretch}{1.1}
\caption{Parameters setup of simulations}
\label{table:parameters}
\centering
\begin{tabular}{lc}
\hline
Parameter & Value \\
\hline\hline
Transmit power ($\mathsf{P}_\mathsf{Tx}$) & $20$~dBm\\
Transmitter antenna gain ($\mathsf{G}_{\rm t}$) & 10~dBi\\
Receiver antenna gain ($\mathsf{G}_{\rm r}$) & 10~dBi\\
Node density of UEs ($\lambda_{\rm UE}$) & [0.0025, 0.005, 0.01] nodes/m$^2$\\
Path-loss exponents of LoS  ($\alpha_{\rm LoS}$) & 2\\
Path-loss exponents of NLoS  ($\alpha_{\rm NLoS}$) & 3\\
Environment & Urban \\
Carrier frequency ($f_{\rm c}$) & $3.5$~GHz \\\hline
\hline
\end{tabular}
\end{table}
\begin{figure}[t!]
	\centering
	\subfloat[Received signal power of the UAV versus the UAV height where $\lambda_{\rm UE}=0.005$ nodes/m$^2$.]{\includegraphics[width=0.48\textwidth]{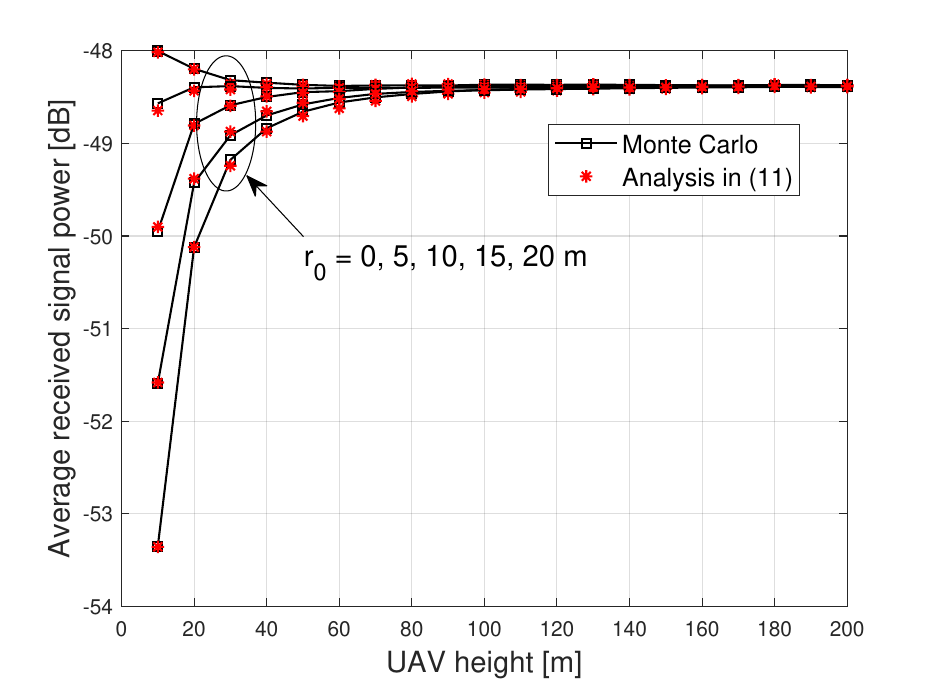}\label{fig:stoch_r0}}
 
        \subfloat[Received signal power of the UAV versus UE node density $\lambda_{\rm UE}$ $r_0 = 10$~m.]{\includegraphics[width=0.48\textwidth]{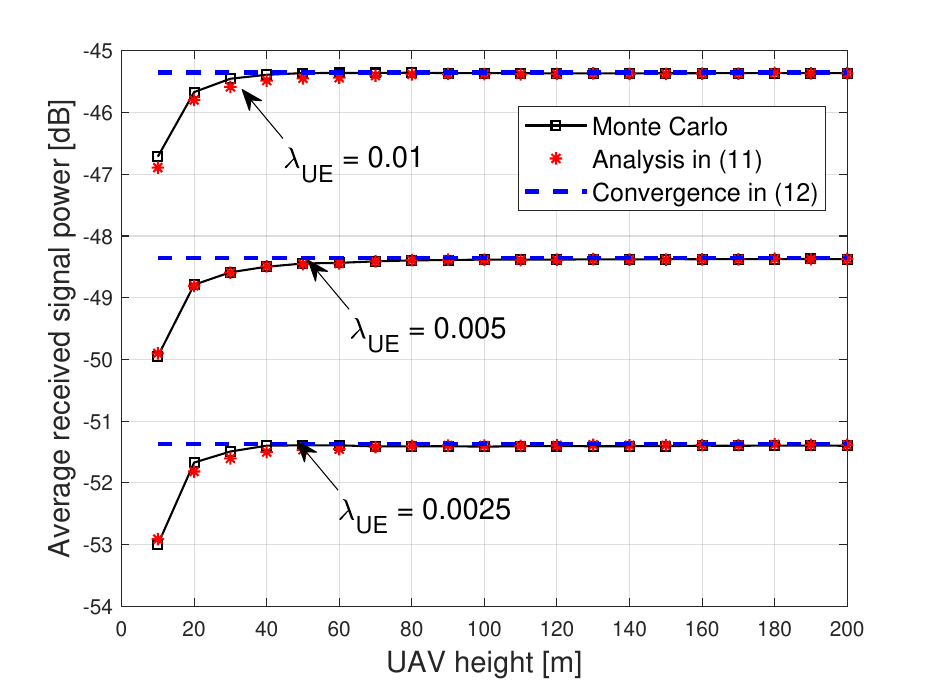}\label{fig:stoch_lambda}}
	\caption{Received signal power of the UAV using the stochastic geometry analysis. Total power is a decreasing function as the UAV height increases when $r_0 = 0$~m, while it becomes an increasing function when $r_0$ is larger than 10~m.}
\end{figure}
\begin{figure}[t!]
	\centering
	\subfloat[LoS probability depending on the 3-D distance with different UAV heights.]{\includegraphics[width=0.48\textwidth]{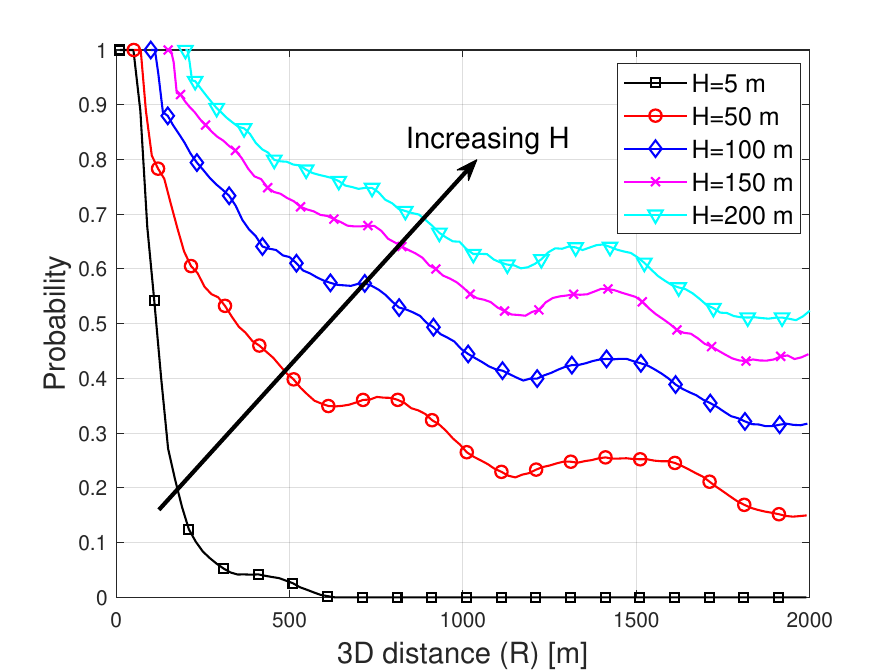}\label{fig:ray_tra_plos}}
 
        \subfloat[Received signal power of the UAV depending on the UAV height.]{\includegraphics[width=0.48\textwidth]{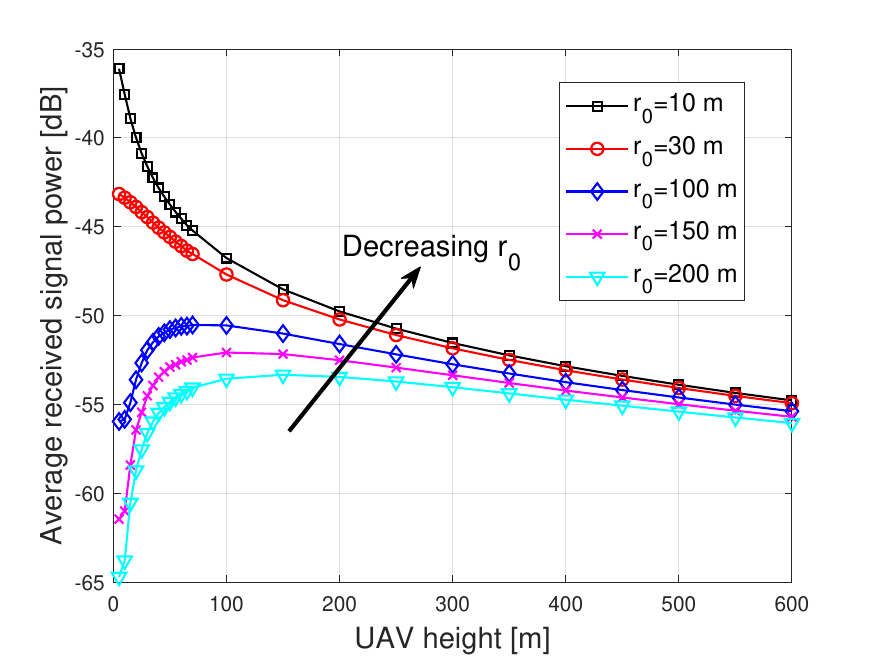}\label{fig:ray_tra_r0}}
	\caption{LoS probability and the aggregate received signal power at the UAV obtained by the ray-tracing analysis model.}
\end{figure}
\begin{figure*}[t!]
	\centering
	\subfloat[UL n71 (663 - 698~MHz).]{\includegraphics[width=0.29\textwidth]{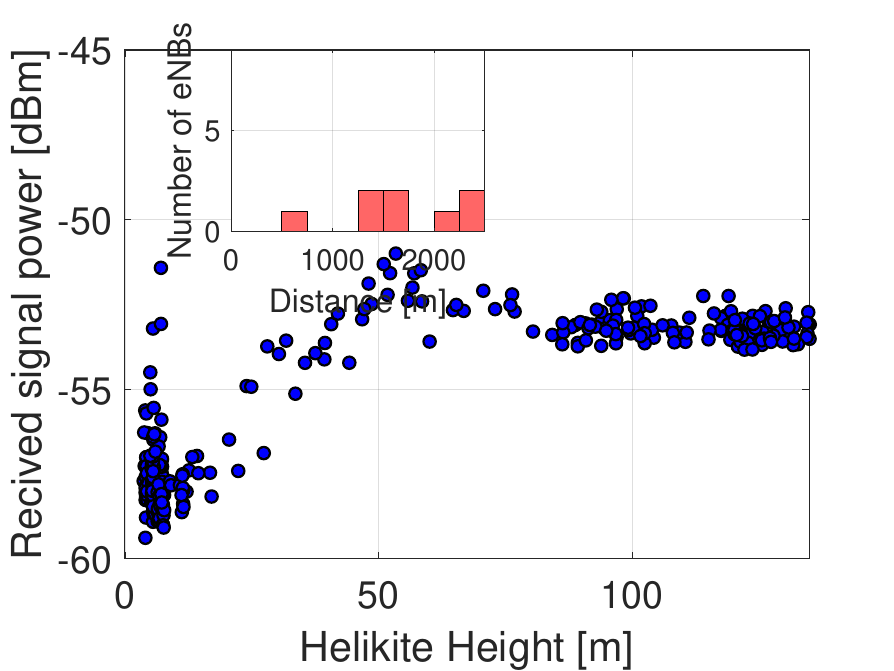}\label{fig:mea_ULn71}}~
        \subfloat[UL 12 (698 - 716~MHz).]{\includegraphics[width=0.29\textwidth]{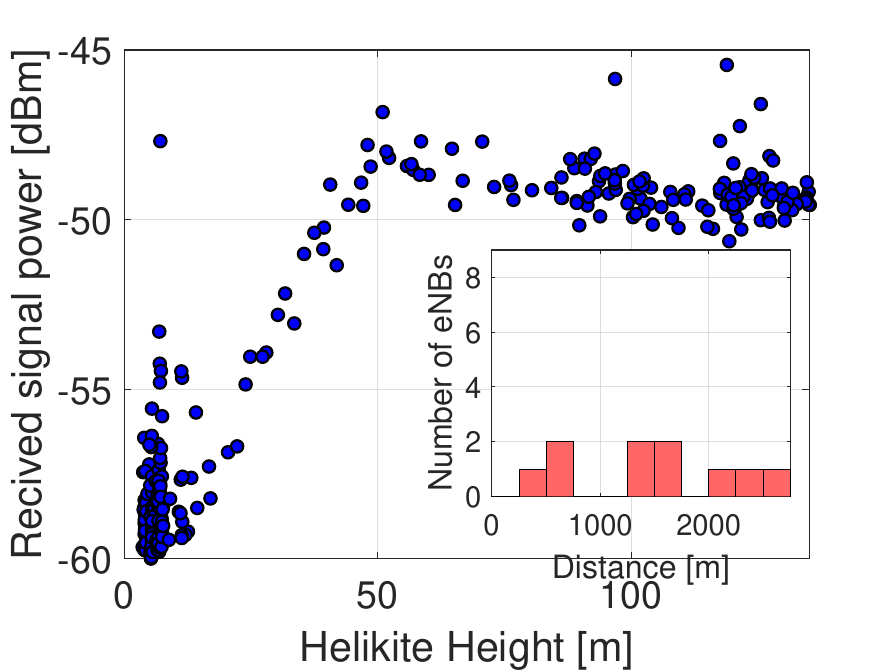}\label{fig:mea_UL12}}~
        \subfloat[UL 5 (824 - 849~MHz).]{\includegraphics[width=0.29\textwidth]{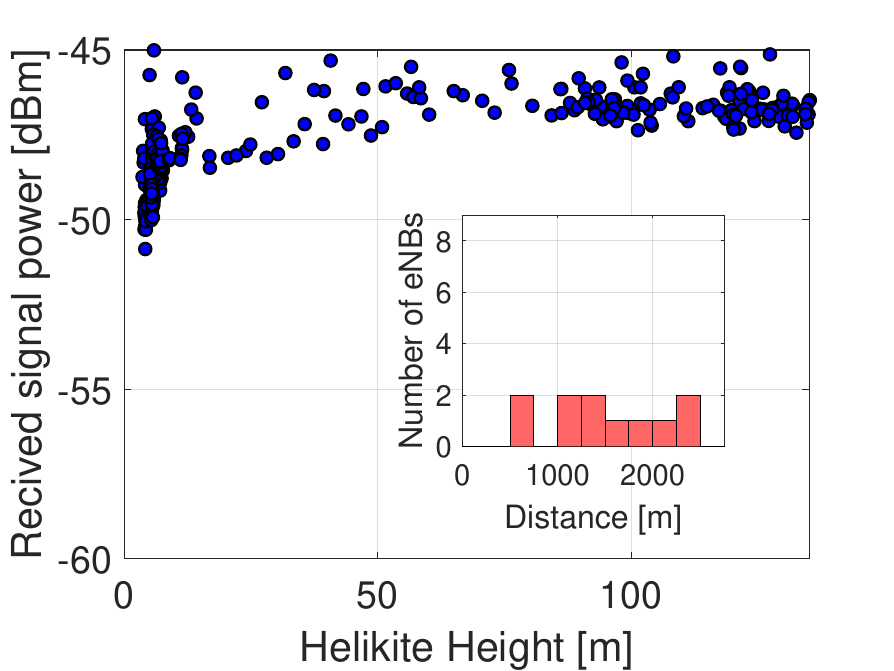}}
        
        \subfloat[UL 4 (1710 - 1755~MHz).]{\includegraphics[width=0.29\textwidth]{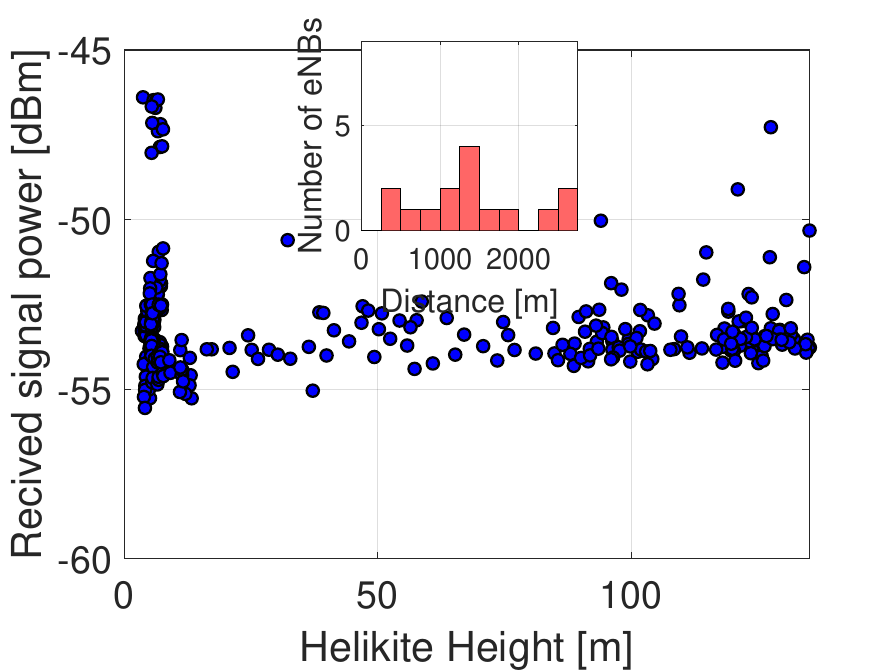}}~
        \subfloat[UL 2 (1850 - 1910~MHz).]{\includegraphics[width=0.29\textwidth]{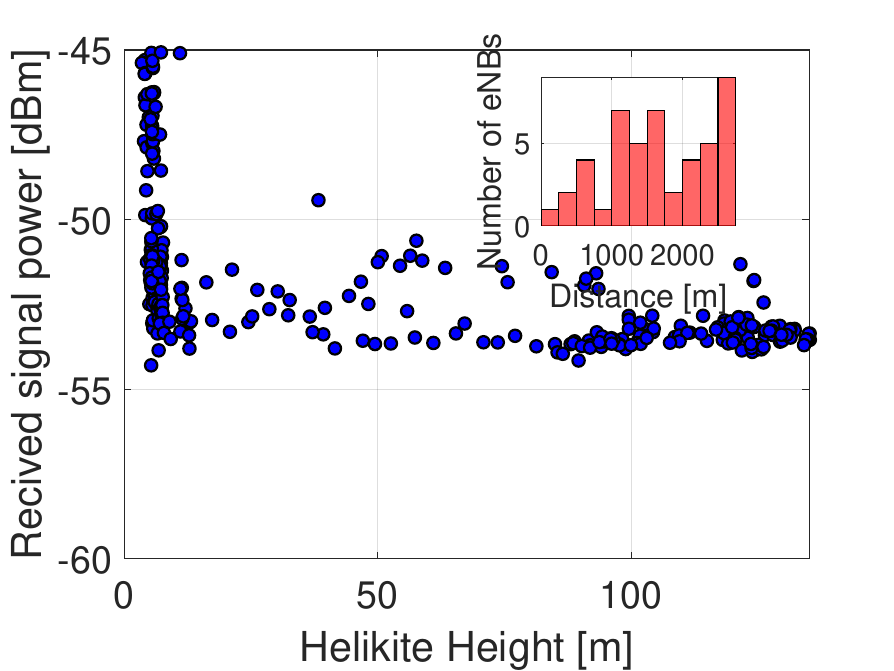}}~
        \subfloat[UL 30 (2305 - 2315~MHz).]{\includegraphics[width=0.29\textwidth]{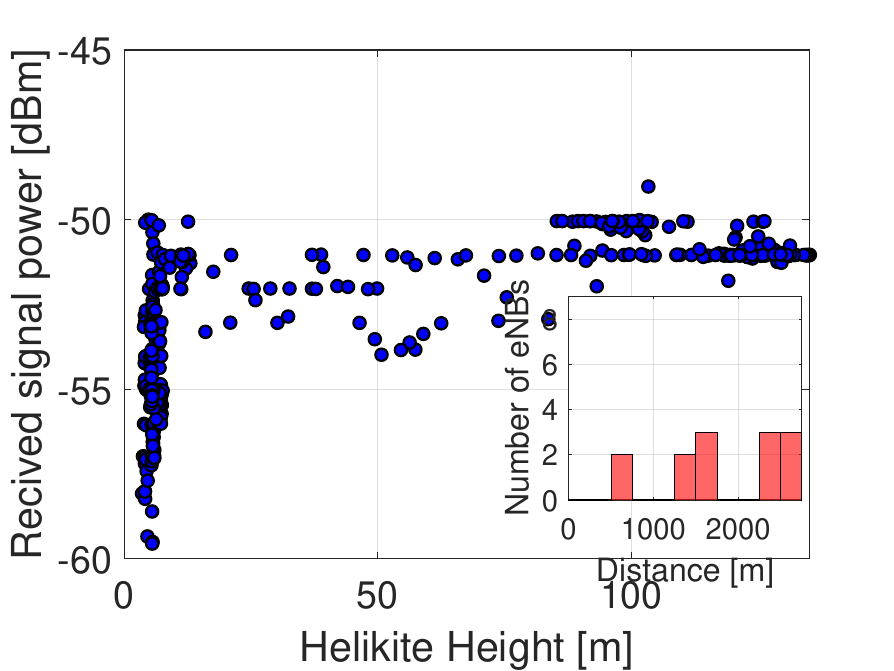}}
	\caption{Received signal power of the helikite during the flight as a function of the helikite height.  The experimental datasets are obtained from the measurement campaign of the spectrum occupancy monitoring during the Packapalooza Festival for the urban environment described in Section~\ref{sec:mea_camp}. The uplink bands are listed in Table~\ref{table:LTE_bands}. The distance distributions of the eNBs and gNBs are also plotted for each UL band. 
 }\label{fig:mea_reci}
\end{figure*}
\begin{figure*}[t!]
	\centering
	\subfloat[UL n71 (663 - 698~MHz).]{\includegraphics[width=0.29\textwidth]{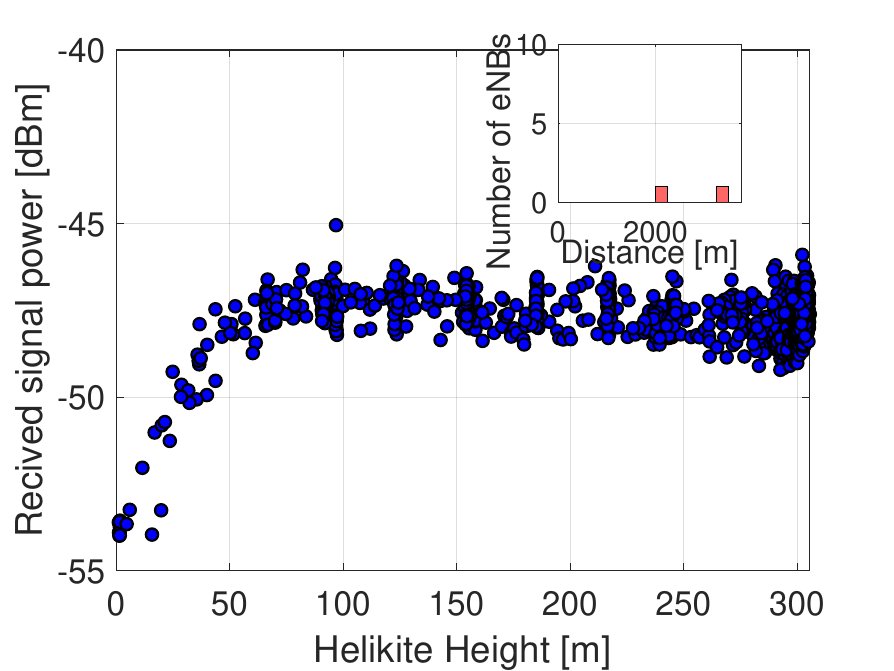}\label{fig:mea_ULn71_LW}}~
        \subfloat[UL 12 (698 - 716~MHz).]{\includegraphics[width=0.29\textwidth]{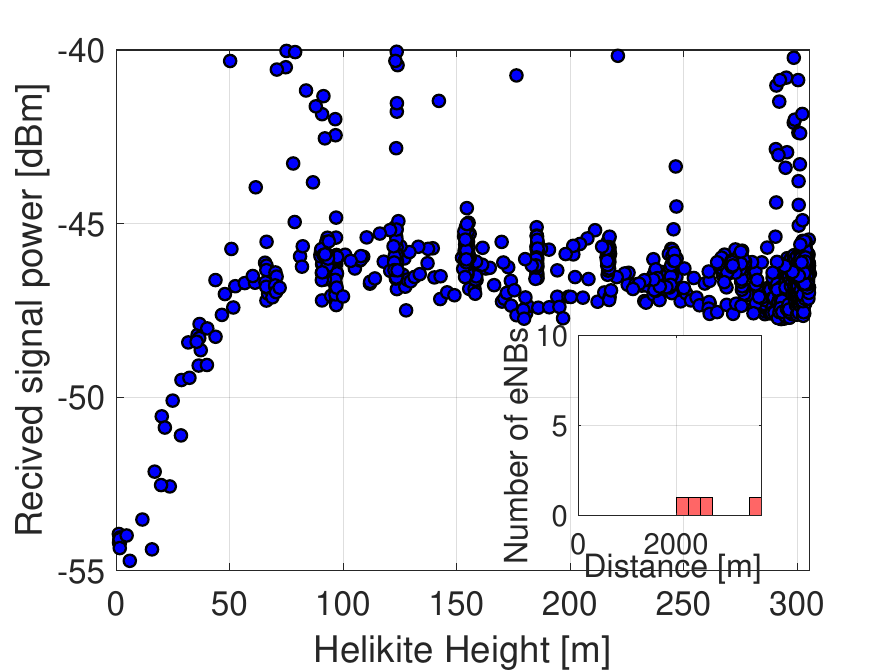}\label{fig:mea_UL12_LW}}~
        \subfloat[UL 5 (824 - 849~MHz).]{\includegraphics[width=0.29\textwidth]{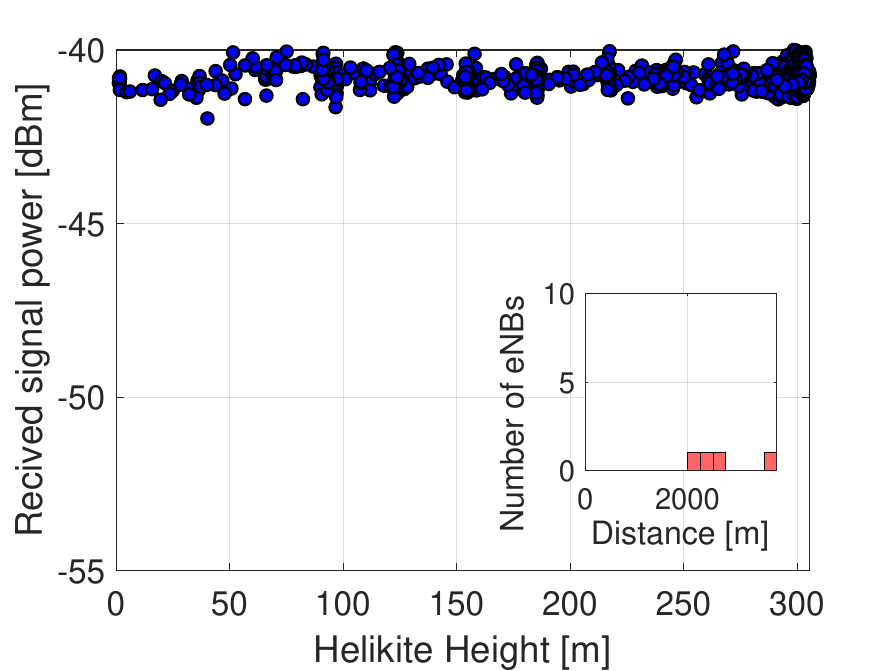}}
        
        \subfloat[UL 4 (1710 - 1755~MHz).]{\includegraphics[width=0.29\textwidth]{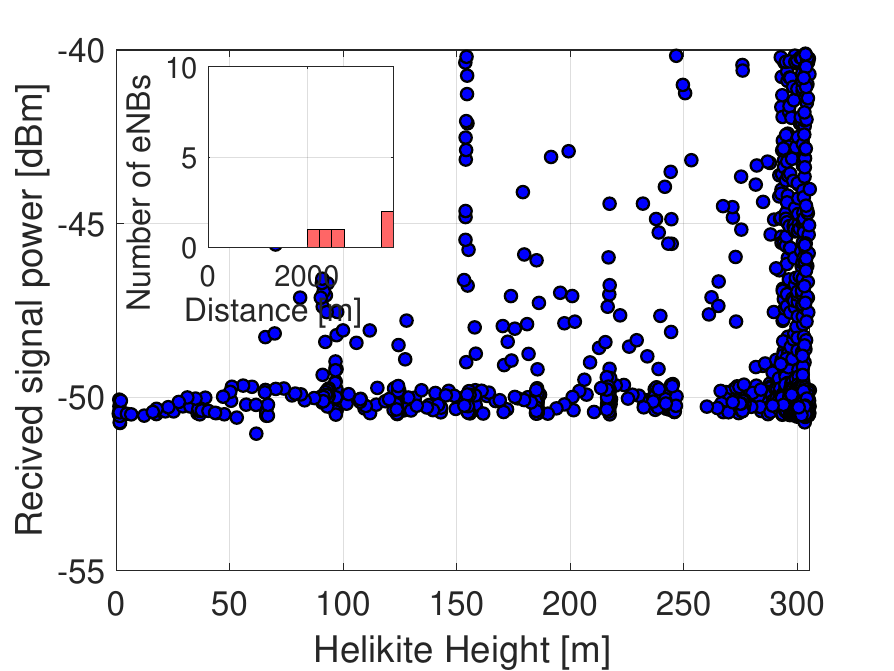}}~
        \subfloat[UL 2 (1850 - 1910~MHz).]{\includegraphics[width=0.29\textwidth]{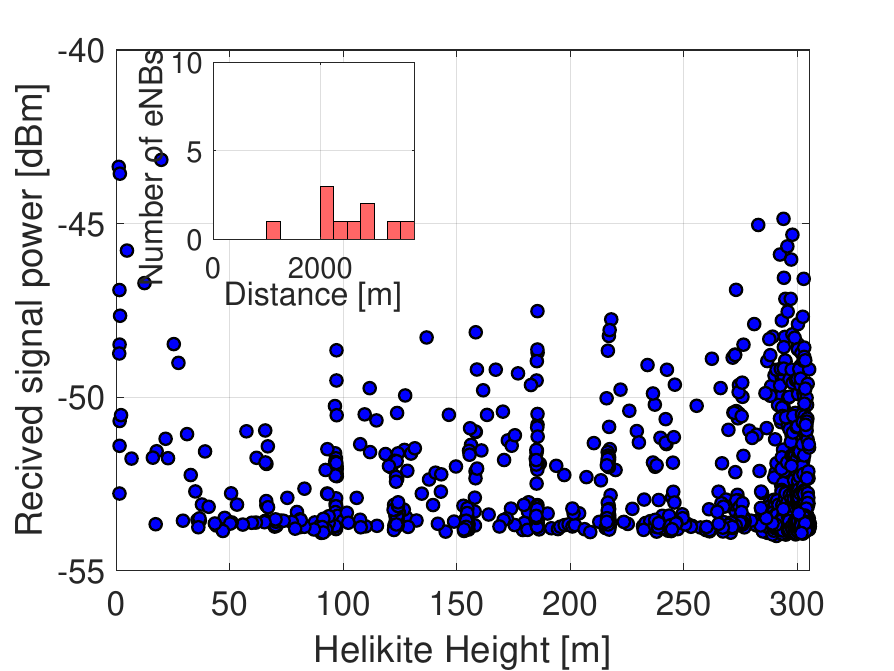}}~
        \subfloat[UL 30 (2305 - 2315~MHz).]{\includegraphics[width=0.29\textwidth]{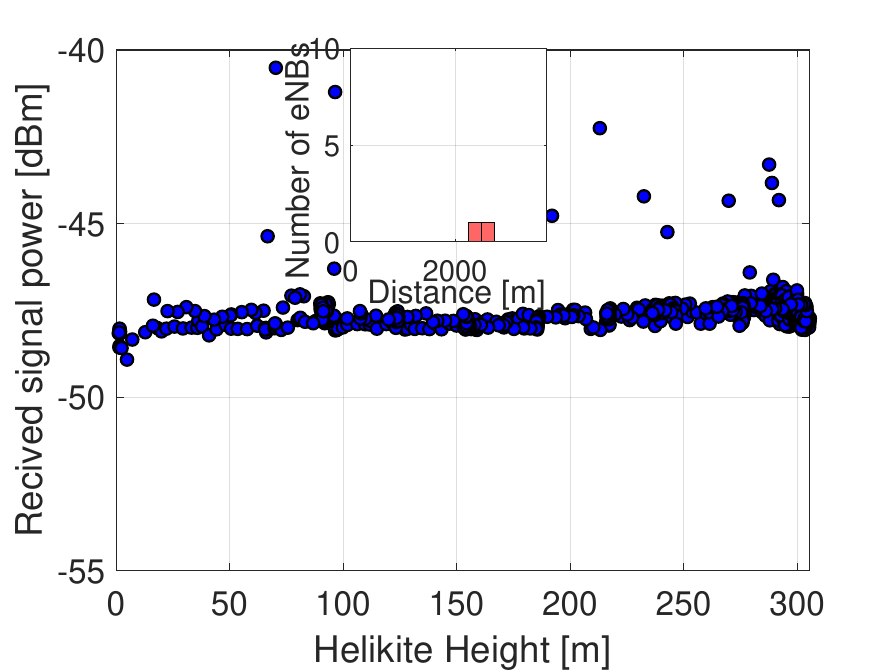}}
	\caption{Received signal power of the helikite during the flight as a function of the helikite height where the experiment is conducted at LakeWheeler field lab for the rural environment. The distance distributions of the eNBs and gNBs are also plotted for each UL band.}\label{fig:mea_reci_lw}
\end{figure*}
\begin{figure}[t!]
	\centering
	\subfloat[Locations of cellular towers of DL 12 obtained from CellMapper.]{\includegraphics[width=0.38\textwidth]{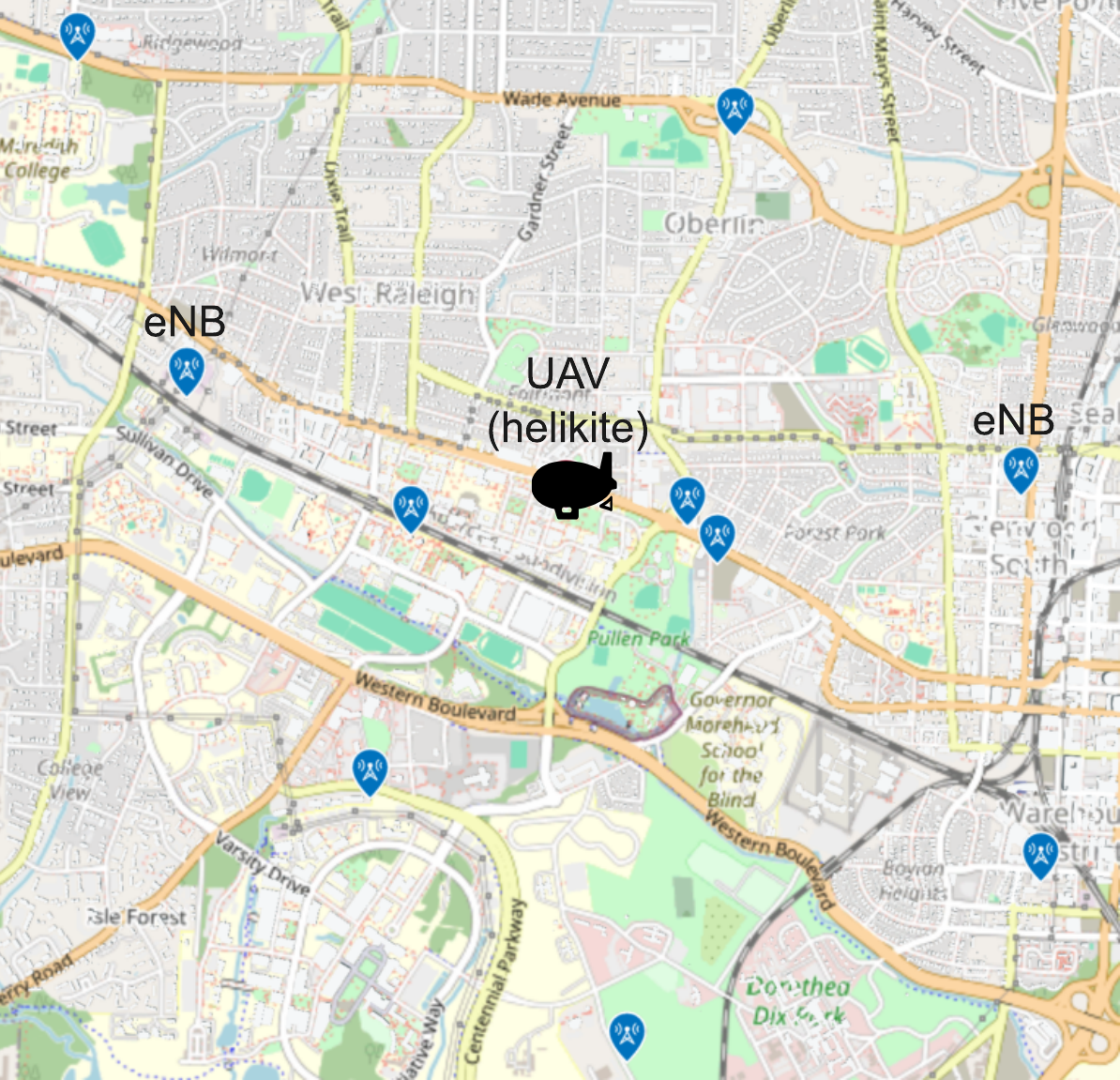}\label{fig:map_DL12}}

        \subfloat[Received signal power of the UAV using the stochastic geometry model, ray-tracing, and measurement.]{\includegraphics[width=0.48\textwidth]{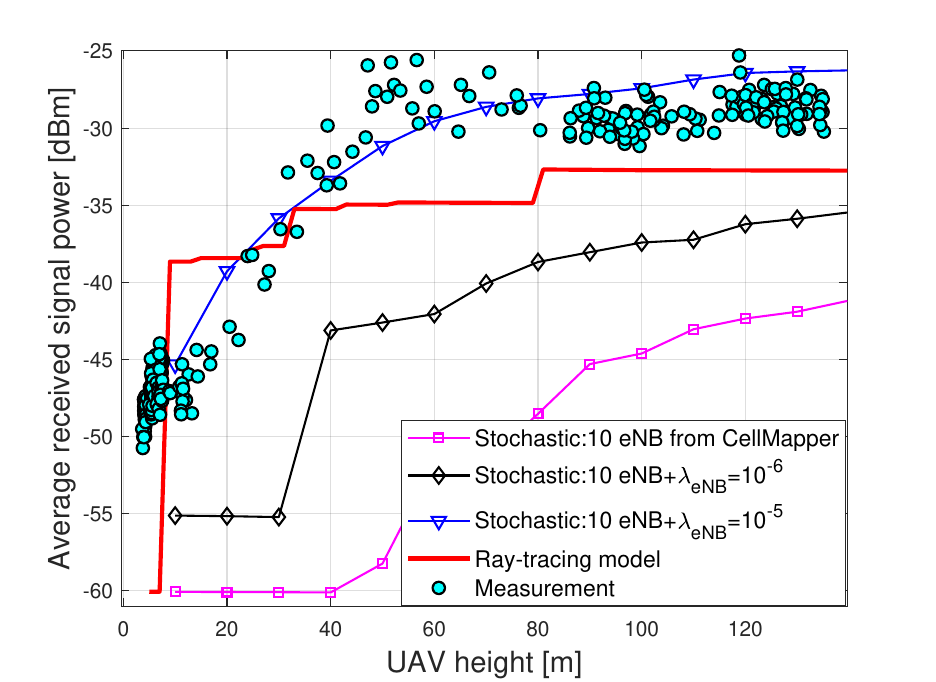}\label{fig:DL12_all}}
	\caption{Downlink spectrum occupancy using  SOSGAD model, ray-tracing, and measurements of DL band 12, where the location of cellular towers are obtained from CellMapper~\cite{cellmapper}.}\label{fig:DL_12_tot}
\end{figure}

First, we provide theoretical results using the SOSGAD model introduced in Section~\ref{sec:stoc_geo}. We consider a scenario where a UAV measures the spectrum occupancy of either UL signals from UEs or DL signals from eNBs and gNBs. The parameters considered in the analytical results are summarized in Table~\ref{table:parameters}. Fig.~\ref{fig:stoch_r0} and Fig.~\ref{fig:stoch_lambda} show the received signal power of the UAV as the UAV height increases. Those results are obtained by the SOSGAD model introduced in Section~\ref{sec:stoc_geo} as well as computer simulations. 
For the Monte Carlo results, we randomly generate the number of UEs following Poisson distribution based on the UE node density $\lambda_{\rm UE}$, and UEs are randomly placed in the 2-D ring-shape area where the outer and inner radius of the ring are 4000~m and $r_0$. The UAV is placed at the xy-origin where the z-axis coordinate is the UAV height $H$. Then, we calculate the aggregate received signal power in \eqref{eq:rece_0} considering the LoS/NLoS link status by using the LoS probability in \eqref{eq:PLoS_ori} for the given experiment area. 

In Fig.~\ref{fig:stoch_r0}, we observe that the analysis from the closed-form expression in \eqref{eq:rece_4} overlaps closely with the Monte Carlo simulation results. It is also observed that when $r_0=0$~m, the received signal power is a decreasing function, while the received signal power is an increasing function when $r_0$ is larger than $5$~m. This observation can be explained by the closed-from expression analysis discussed in Section~\ref{sec:lammda_r0}. In addition, according to \textit{Theorem}~1, the received signal power converges to the same value for all $r_0$ values.

In Fig.~\ref{fig:stoch_lambda}, we show the impact of the UE node density $\lambda_{\rm UE}$ on the received signal power. As we expect from the analysis in Section~\ref{sec:lammda_r0}, as the UE node density increases from $0.0025$ to $0.005$ and finally to $0.01$, the received signal power increases linearly. 
Furthermore, we can observe that the asymptotic value of the received signal power in \textit{Theorem}~1 is close to the convergence value of the received signal power as the UAV height increases.

\subsection{Spectrum Occupancy Results from Ray-Tracing Analysis}
Fig.~\ref{fig:ray_tra_plos} shows the LoS probability as a function of the 3-D distance $R$ for varying UAV height $H$. The results are generated by the ray-tracing analysis of the LoS/NLoS status of all the UEs on the grid as described in Fig.~\ref{fig:LoS_ray} in Section~\ref{sec:ray_tracing}. 

When compared with the statistically formulated expression of the LoS probability function in the urban environment in Fig.~\ref{fig:plos_1}, we observe that the LoS probability reduces slowly and does not become zero in Fig.~\ref{fig:ray_tra_plos}. This is because the considered real-world environment around NC State Main Campus is not fully surrounded by buildings. The south side includes a relatively rural area and there are large green fields with no buildings. On the other hand, the north side of the area is relatively urban with more buildings. For this reason, there are many UEs in the south direction where the LoS condition is satisfied even when located at a far distance, which can also be observed in Fig.~\ref{fig:LoS_ray}. From the results, we can expect that the LoS probability function in the real-world can deviate in practice from the statistical function in~\eqref{eq:PLoS_ori} depending on the distribution of the buildings.

Fig.~\ref{fig:ray_tra_r0} shows the received signal power based on ray tracing analysis as a function of the UAV height with different $r_0$. When we compare it with Fig.~\ref{fig:stoch_r0} obtained from the SOSGAD model, we observe that the trends of the convergence curves and the impact of $r_0$ 
for different UAV heights 
are similar. However, due to the different trends of the LoS probability function in Fig.~\ref{fig:ray_tra_plos} compared with Fig.~\ref{fig:plos_1}, the convergence becomes much slower compared with the stochastic geometry model in Fig.~\ref{fig:stoch_r0}. In addition, the received signal power becomes a decreasing function with a larger $r_0$.

\subsection{Spectrum Occupancy Results from Measurements}
In Fig.~\ref{fig:mea_reci}, we show the spectrum occupancy monitoring results in the urban area obtained using a helikite as discussed in Section~\ref{sec:mea_camp}. We consider the uplink frequency bands of LTE and 5G NR cellular networks listed in Table~\ref{table:LTE_bands}. During the helikite flight, 364 spectrum sweeps are carried out, and the received signal power versus the helikite height are indicated by the blue dots in the figures. In addition, we show the distance distribution of the eNBs for each band, where the locations of the eNBs are obtained from CellMapper~\cite{cellmapper}.

We observe that the received signal power gradually increases and becomes constant in UL n71, 12, 5, and 30. This implies that received signal power increases as more UEs are linked by LoS conditions, which is observed in the stochastic geometry model and the ray-tracing model. On the other hand, the received signal is relatively constant in UL 4 and 2 with high peak power at the low height. The high peak and the large variation of the signal power at the low height could come from the multi-path effects, antenna orientation change at UEs and the helikite, power control of UEs in the uplink, and the high density of UEs during certain time intervals. There may also UEs very close to the helikite that may be scheduled intermittently in the uplink, causing a large variance in the total received power. 

In addition, it is observed that the received signal power slowly increases in UL n71, 12 compared with UL 5, 30. This implies that depending on the geographical distribution of the UEs using bands, the slope of received signal power increase with altitude can vary. Lastly, when we compare the measurement results with ray-tracing results in Fig.~\ref{fig:ray_tra_r0}, the height of the UAV when the received signal power converges to a fixed value is much lower in measurements compared with the ray-tracing model. This is because the UEs are densely populated on the streets during the experiment while the festival was ongoing, while the UEs are uniformly placed on the area in the ray-tracing model.

Fig.~\ref{fig:mea_reci_lw} shows the spectrum occupancy monitoring results from the experiment in the rural area. Similar to the urban area results, it is observed that the received signal gradually increases and becomes constant in UL bands n71, 12. However, the increasing slopes are relatively slow and hold constant values up to the maximum height of around 300~m, which is around 140~m in the urban scenario. We also observe that the constant value of the received signal starts from the low height in UL bands 5, 4, 2, and 30. In addition, the high peak power at the height of around 300~m is observed in UL bands 4 and 2, which comes from similar reasons to the urban results in the low height since the helikite stayed at the same height for a long time (around 4 hours).

\subsection{Stocahstic Geometry, Ray-tracing, and Measurement Results for Downlink Spectrum Occupancy}
In the earlier subsections, we focus on the uplink signals from UEs distributed on the ground. In this subsection, we additionally study the received signal power of the downlink (DL) signal. Unlike uplink signals from users, the locations of the cellular towers (eNBs and gNBs) can be found publicly. As an example, we show the locations of eNBs of DL~12 (728 - 746~MHz) within 2.75~km of the measurement in Fig.~\ref{fig:map_DL12} based on the information at CellMapper~\cite{cellmapper}. Since a total of ten eNBs from CellMapper may not represent all eNBs near the measurement site, there would be more eNBs that are not listed in the crowd-sourced data. 
Fig.~\ref{fig:DL12_all} compares the received signal power of the UAV generated by the stochastic geometry model, the ray-tracing model, and measurement data. The result of the ray-tracing model shows the total received signal power at the UAV based on the locations of ten eNBs in Fig.~\ref{fig:map_DL12}. We observe that as the UAV height increases, more eNBs observe a LoS condition, and the received signal power increases. After most of the eNBs are in the LoS condition, the received signal power converges to a constant value.

In the stochastic geometry model in Fig.~\ref{fig:DL12_all}, after calculating the distance between the helikite and 10 eNBs in Fig.~\ref{fig:map_DL12} and fixing the location of 10 eNBs, we randomly deploy additional eNBs following the node density $\lambda_{\rm eNB}$. The square marker curve indicates the case that we only consider fixed 10 eNBs from CellMapper without additional eNBs. In that case, it is observed that the average received signal starts to increase from the UAV height of around 40~m. In addition, as the randomly deployed eNBs are added, the average received signal power becomes a smoothly increasing curve.

The measurement result in Fig.~\ref{fig:DL12_all} shows the experimental results from the measurement campaign of the received signal in the DL band 12. We observe that the received signal gradually increases and becomes constant, which has a similar tendency as in the UL received power in Fig.~\ref{fig:mea_ULn71}, Fig.~\ref{fig:mea_UL12}. The ray-tracing-based received power is seen to underestimate the total received power observed from measurements. We believe this is because not all the eNBs are captured in the CellMapper database, and there are other distant eNBs that may be contributing to the total received power. This could also explain the step-wise behavior of the aggregate received signal power in ray-tracing results. 
The differences in Fig.~\ref{fig:DL12_all} between measurements and analytical models can also be attributed to not using the exact density and locations of BSs, as well as the impacts of multi-path fading, scattering, and specific 3D antenna patterns. 

\section{Conclusion}\label{sec:Conclusion}
In this paper, we study different approaches for modeling the altitude-dependent occupancy of the cellular spectrum of UL/DL in urban and rural environments. We propose the SOSGAD model and analyze the aggregate received signal power as a function of the UAV height. By deriving closed-form expressions, we show that the received signal power converges to a constant value as the UAV height increases. In addition, we closely approximate the altitude-dependence of the LoS probability function as a function of the 3-D distance, which is useful to obtain mathematically tractable expressions. Moreover, we compare the same spectrum occupancy monitoring scenario by using the real-world 3-D map and ray-tracing model for analyzing LoS/NLoS link status. Finally, we compare the results from the analysis by experimental datasets from the spectrum occupancy measurement campaign by using helikite.

\bibliographystyle{IEEEtran}
\bibliography{IEEEabrv,references}
 
\end{document}